\documentclass[preprint,12pt,sort&compress]{elsarticle}

\graphicspath{{./}}
\usepackage{hyperref}
\usepackage{xcolor}
\usepackage{placeins}
\usepackage{duckuments}
\usepackage{amssymb}
\usepackage[T1]{fontenc}    
\usepackage{lmodern}
\usepackage{amsfonts,amsmath}
\usepackage{graphicx}
\usepackage{subcaption}
\usepackage{algorithm}
\usepackage{algorithmicx}
\usepackage[normalem]{ulem}

\journal{Journal of the Mechanics and Physics of Solids}

\begin{document}

\begin{titlepage}
	\color[rgb]{.4,.4,1}
	\hspace{5mm}

	\bigskip
	
	\hspace{15mm}
	\begin{minipage}{133mm}
		\color{black}
		\sffamily
		\LARGE\bfseries Recovering Mullins damage hyperelastic behaviour with physics augmented neural networks \\[-0.3\baselineskip] 
		
		\vspace{5mm}
		{\large {Preprint of the article published in \\[-0.4\baselineskip] Journal of the Mechanics and Physics of Solids (2024) }} 
		
		\vspace{10mm}        
		{\large Martin Zlati\'{c}, Marko \v{C}ana\dj{}ija } 
		
		\large
		
		\vspace{40mm}
		\vspace{5mm}
		
		\small
		\url{https://doi.org/10.1016/j.jmps.2024.105839}
		
		\textcircled{c} 2024. This manuscript version is made available under the CC-BY-NC-ND 4.0 license \url{http://creativecommons.org/licenses/by-nc-nd/4.0/}
		\hspace{30mm} 
		\color[rgb]{.4,.4,1} 
		\includegraphics[width=3cm]{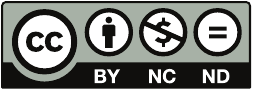}        
	\end{minipage}
\end{titlepage}

\begin{frontmatter}



\title{Recovering Mullins damage hyperelastic behaviour with physics augmented neural networks}

\author[label1]{Martin Zlati\'{c}}
\author[label1]{Marko \v{C}ana\dj ija\corref{cor}}
\affiliation[label1]{organization={Faculty of Engineering, University of Rijeka},
             addressline={Vukovarska 58},
             city={Rijeka},
             postcode={51000},
             country={Croatia}}
\cortext[cor]{Corresponding author}

\begin{abstract}

The aim of this work is to develop a neural network for modelling incompressible hyperelastic behaviour with isotropic damage, the so-called Mullins effect. This is obtained through the use of feed-forward neural networks with special attention to the architecture of the network in order to fulfil several physical restrictions such as objectivity, polyconvexity, non-negativity, material symmetry and thermodynamic consistency. The result is a compact neural network with few parameters that is able to reconstruct the hyperelastic behaviour with Mullins-type damage. The network is trained with artificially generated plane stress data and even correctly captures the full 3D behaviour with much more complex loading conditions. The energy and stress responses are correctly captured, as well as the evolution of the damage. The resulting neural network can be seamlessly implemented in widely used simulation software. Implementation details are provided and all numerical examples are performed in Abaqus.


\end{abstract}




\end{frontmatter}


\section{Introduction}\label{sec:intro}

The motivation behind using neural networks (NNs) for describing material behaviour is to ultimately replace conventional models with new, generalised ones. There are 2 types of approaches, those that work on data alone and machine learning models trained on data. The first ones completely do away with models, marking them as model-free approaches and "true" data-driven techniques using data obtained either from experiments or simulations. This was first introduced in \cite{Kirchdoerfer2016} as data-driven computational mechanics. It has been extended to hyperelasticity \cite{Platzer2021}, inelastic behaviour \cite{Eggersmann2019}, molecular dynamics simulations \cite{Bulin2023}, multiscale modelling \cite{Karapiperis2021} and other areas. The second type of approaches is based around machine learning models, of which this work makes use of. More specifically, neural network models are used to capture the Mullins effect of incompressible hyperelastic materials, observed in \cite{Mullins1948}. The use of neural networks in material modelling is now commonplace, specifically a special type known as \textit{feed-forward neural networks} (FNNs). The usual concept of using neural networks as regression operators is no longer sufficient, but further refinements must be added. It has been known for some time that NNs can be used as universal approximators \cite{Hornik1989} and as black-box material models. A very early example of this can be found in \cite{Ghaboussi1991}, where the behaviour of concrete under cyclic loading was described. In an attempt to describe hyperelastic behaviour, an NN was used in \cite{Shen2004} to predict strain energy from invariants, or in \cite{Weber2021} the neural network was used to predict stresses from strains. Both approaches were used to describe adiabatic thermo-hyperelasticity in \cite{Zlatic2023}. In \cite{Huang2020}, the use of NNs in conjunction with proper orthogonal decomposition is used to model von Mises plasticity in addition to hyperelastic behaviour. Related to this work, in \cite{Ghaderi2020} multiple neural networks are used in a multi-agent system to model the Mullins effect, and is to the authors knowledge the first work to treat the Mullins effect with machine learning. In \cite{Abdusalamov2024}, deep symbolic regression is used to recover the Mullins effect on several types of rubber behaviour (compressible, incompressible, and temperature dependent). The Mullins effect is exhibited by many hyperelastic materials that manifests as strain induced stress softening. In this work is described as an isotropic damage model that is added to the primary material response via a scalar damage parameter. A more detailed overview of the Mullins effect is given later in the work.

As the field progresses certain refinements are included. Namely, thermodynamic consistency, objectivity, normalisation and non-negativity of the energy and convexity. Thermodynamic consistency is fulfilled by the use of suitable inputs and outputs \cite{Zlatic2023,Linka2021,Thakolkaran2022,Klein2022,Linden2023}, by the complete reconstruction of the NN architecture \cite{Masi2021}, or it can be treated \textit{a posteriori} \cite{Kalina2021}. The preservation of convexity is a constraint that has been implemented in various forms. In \cite{amos17}, the authors proposed a new form of NNs called Input Convex Neural Networks (ICNN), where the neural network is a convex function with respect to its inputs. This was adopted by \cite{Thakolkaran2022} to model hyperelastic behaviour by predicting strain energy based on invariants. In \cite{Klein2022,Linden2023,Klein2023} the condition of polyconvexity is enforced by a suitable choice of input invariants and activation function as well as by restricting certain weighting factors to be non-negative. Objectivity can be fulfilled by training the NN on many examples including rotated states of stress-strain, \cite{Weber2021,Zlatic2023,Klein2022}, or by employing invariants \cite{Linka2021,Kalina2021,Klein2022,Thakolkaran2022,Zlatic2023}. Employing invariants offers a considerable advantage in that much less training examples are needed and should be considered a more preferable option. Normalisation of the energy can be done in two ways, exactly or approximately. Exactly by choosing an appropriate activation function and input so that the energy in an undeformed state is equal to zero. This could be as simple as using the hyperbolic tangent or an exponential linear unit which are centred at zero and setting the inputs as such that the output is zero in the undeformed state \cite{Zlatic2023,Kalina2021,Linka2021}. Approximately by training the network with augmented data containing the undeformed state. Non-negativity of the strain energy can be enforced directly through the use of suitable activation functions \cite{Klein2022,Linden2023} and limiting certain parameters to be non-negative. Furthermore, adopting an NN that predicts energy but is trained on stress data would mean training the NN on its derivatives which is refferred to as Sobolev traning and is done in the works of \cite{Kalina2024,Vlassis2020,Vlassis2021,Czarnecki2017,Klein2023}.

This work presents a novel NN architecture following the constraints of previous works on physics augmented neural networks and modelling techniques from constitutive artificial neural networks adopted to include Mullins type isotropic damage. The result is an NN model of modest size and high accuracy, trained on data obtainable from simple tests, with applications on complex loading conditions.

\section{Hyperelasticity and the Mullins effect}\label{sec:Mullins_effect}

Incompressible hyperelastic behaviour with Mullins effect is considered in this paper. Hyperelasticity is modelled with a suitable strain energy function $\psi$ based on the principal stretches of the right $(\mathbf{U})$ or left $(\mathbf{V})$ stretch tensor or on the invariants of the right ($\mathbf{C}$) or left ($\mathbf{B}$) Cauchy-Green deformation tensor. A strain energy function is derived with respect to a chosen deformation measure (the deformation gradient $\mathbf{F}$ or the tensors $\mathbf{C}$ or $\mathbf{B}$), depending on the stress measures we want to derive. The Clausius-Duhem equation can be reduced in the case of a hyperelastic material to the following:
\begin{equation}
	\mathbf{P}:\dot{\mathbf{F}}-\dot{\hat{\psi}}\geq 0 \rightarrow \left(\mathbf{P} - \frac{\partial{\hat{\psi}(\mathbf{F})}}{\partial{\mathbf{F}}}\right):\dot{\mathbf{F}} \geq 0,
	\label{eq:Clausius_Planck_ineq}
\end{equation}
where the $1^\text{st}$ Piola-Kirchhoff stress tensor $\mathbf{P}$ is defined in terms of the derivative of the strain energy $\psi$ with respect to the deformation gradient $\mathbf{F}$. The general expression for the strain energy in case of incompressible behaviour ($J=\text{det}\mathbf{F}=1$) can be given as 
\begin{equation}
	\hat{\psi}(\mathbf{F}) = \psi(\mathbf{F}) - p(J-1),
\end{equation}
where $p$ is an indeterminate Lagrange multiplier added in the case of incompressibility. It is a workless constraint that has no contribution to the energy, but when deriving the strain energy with respect to $\mathbf{F}$ the expression for the 1$^\text{st}$ Piola-Kirchhoff stress is obtained as
\begin{equation}
	\mathbf{P} = -p\mathbf{F}^{-\text{T}} + \frac{\partial{\psi(\mathbf{F})}}{\partial{\mathbf{F}}}.
	\label{eq:1PK_tensor_form}
\end{equation}
The symbol $\psi$ denotes a function that defines the strain energy that occurs purely from change in shape , i.e. isochoric deformation. The strain energy function is most often not used directly as a function of $\mathbf{F}$. As mentioned, there are two most common ways of defining the strain energy function, either by using the eigenvalues (principal stretches) of $\mathbf{U}$ and $\mathbf{V}$, or the invariants of $\mathbf{C}$ and $\mathbf{B}$. Both ways are used in this work, the first as an analytical model from which data is generated and the results compared against, and the other as the approach on which the NN is based. In terms of principal stretches, the strain energy $\psi$ can be defined using the Ogden model which is used as a reference for the generation of training data and the error calculation. The 3-term model was adopted from \cite{Ogden1972} in the following form
\begin{equation}
	\psi_\text{Ogden}(\lambda_1,\lambda_2,\lambda_3) = \sum_{p = 1}^{3} \frac{\mu_p}{\alpha_p}\left(\lambda_1^{\alpha_p} + \lambda_2^{\alpha_p} + \lambda_3^{\alpha_p} - 3\right),
	\label{eq:Ogden_model}
\end{equation}
where $\mu_p$ are the shear moduli, $\alpha_p$ are dimensionless constants, see Table \ref{tab:ogden_constants}, and $\lambda_1,\lambda_2,\lambda_3$ are the principal stretches of $\mathbf{U}$ or $\mathbf{V}$. Stress $\mathbf{P}$ can be expressed in terms of principal stretches in the following form
\begin{equation}
	P_a = -p\frac{1}{\lambda_a} + \frac{\partial{\psi_\text{Ogden}(\lambda_1,\lambda_2,\lambda_3)}}{\partial{\lambda_a}},\; \mathbf{P} = \sum_{a = 1}^{3}P_a\mathbf{n}_a\otimes\mathbf{N}_a,
	\label{eq:1PK_definition}
\end{equation}
where $P_a$ is the principal value of the $1^\text{st}$ Piola-Kirchhoff stress tensor ($a = 1,2,3$), $\mathbf{n}_a$ are the eigenvectors of the left stretch tensor $\mathbf{V}$, and $\mathbf{N}_a$ are the eigenvectors of the right stretch tensor $\mathbf{U}$.
\begin{table}
	\centering
	\begin{tabular}{ c|c|c }
		$p$ & $\mu_p$ [MPa] & $\alpha_p$ [-]\\
		\hline
		1 & 0.63 & 1.3 \\
		\hline
		2 & 0.0012 & 5 \\
		\hline
		3 & -0.01 & -2 \\
		
	\end{tabular}
	\caption{Material constants for the Ogden model \cite{Ogden1972}.}
	\label{tab:ogden_constants}
\end{table}

A definition of the $2^\text{nd}$ Piola-Kirchhoff stress tensor $\mathbf{S}$ in terms of the invariants of the right Cauchy-Green deformation tensor $\mathbf{C}$ is obtained by differentiating a strain energy function $\hat{\psi}$ with respect to $\mathbf{C}$ as
\begin{equation}
	\mathbf{S} = 2\frac{\partial{\psi(I_1,I_2)}}{\partial{\mathbf{C}}} - p\mathbf{C}^{-1} = 2\left(\frac{\partial{\psi}}{\partial{I_1}} + I_1\frac{\partial{\psi}}{\partial{I_2}}\right)\mathbf{I} - 2\frac{\partial{\psi}}{\partial{I_2}}\mathbf{C} - p\mathbf{C}^{-1},
	\label{eq:2PK_definition}
\end{equation}
where $\psi(I_1,I_2)$ is a strain energy as a function of invariants, $I_1$ and $I_2$ are the invariants of $\mathbf{C} = \mathbf{F}^\text{T}\mathbf{F}$ which are calculated as $I_1 = \text{tr}(\mathbf{C})$ $= ||\mathbf{F}||^2$ and $I_2 = \frac{1}{2}\left[\left(\text{tr}(\mathbf{C})\right)^2 - \text{tr}\left(\mathbf{C}^2\right) \right]=||\text{Cof}\; \mathbf{F}||^2$ , and $p$ is the same Lagrange multiplier previously encountered in Eq.~(\ref{eq:1PK_definition}). The Lagrange multiplier $p$ in Eq.~(\ref{eq:2PK_definition}) is obtained from equilibrium equations and boundary conditions for a particular problem. For example, it can be expressed from the plane stress condition $S_3 = 0$ by spectral decomposition of the $2^\text{nd}$ Piola-Kirchhoff stress tensor. The resulting expression is 
\begin{equation}
	p = 2\lambda_3^2\left[\frac{\partial{\psi}}{\partial{I_1}} + \left(\lambda_1^2 + \lambda_2^2\right)\frac{\partial{\psi}}{\partial{I_2}}\right],
	\label{eq:p_definition}
\end{equation}
and is given here for completeness. This expression will also be necessary during NN training. Finally, the relation between $\mathbf{P}$ and $\mathbf{S}$ is given by $\mathbf{S} = \mathbf{F}^{-1}\mathbf{P}$, and using the property $(\frac{\partial{\psi(\mathbf{F})}}{\partial{\mathbf{F}}})^\text{T} = 2\frac{\partial{\psi(\mathbf{C})}}{\partial{\mathbf{C}}}\mathbf{F}^\text{T}$ one can obtain Eq.~(\ref{eq:2PK_definition}) from Eq.~(\ref{eq:1PK_definition}) and vice versa. At this point it is remarked that the incompressible Ogden model used above is a polyconvex model, i.e. $\psi_\text{Ogden} = \psi_\text{Ogden}(\mathbf{F},\text{Cof}\;\mathbf{F})$, meaning it has to be convex with respect to $\mathbf{F},\text{Cof}\;\mathbf{F}$. The proof is provided in the landmark papers \cite{ball1976convexity,ball1977constitutive}.

Elasticity tensors are calculated by further deriving the stresses with respect to the chosen deformation measure. If the elasticity tensor is required in the material description, then the $2^\text{nd}$ Piola-Kirchhoff tensor is further derived with respect to $\mathbf{C}$, $\mathbb{C} = 2\frac{\partial{\mathbf{S}(\mathbf{C})}}{\partial{\mathbf{C}}}$. For the calculation of the stress and elasticity tensors in both the principle stretch and the invariant formulation, the reader is referred to \cite{Connolly2019}, where the numerical implementation is also described in detail.

A common example of an invariant-based model is the Mooney-Rivlin model with the form
\begin{equation}
	\psi_\text{MR}(I_1, I_2) = C_{10}(I_1-3) + C_{01}(I_2-3),
	\label{eq:Mooney-Rivlin}
\end{equation}
where $C_{10}$ and $C_{01}$ are material constants. An overview of many different models was given by \cite{Steinmann2012}. The models based on invariants can be easily derived to obtain stress and elasticity tensors and are easy to implement in common industrial FEA software through dedicated subroutines.

When modelling material behaviour, it is desirable to include certain constraints to the material model. In this work, these are the included constraints:
\begin{enumerate}
	\item objectivity
	\item normalisation condition of the energy
	\item normalisation condition of the stress
	\item non-negative strain energy
	\item convexity
	\item thermodynamic consistency
\end{enumerate}

These constraints have also been included in other works, e.g. \cite{Linden2023,Linka2023,Kalina2024}, and their fulfilment is a step towards building a fully functional NN model that respects conditions from solid mechanics. More will be said on the inclusion of these constraints in Section~\ref{sec:NN_arch}. The most notable one is \textit{objectivity} which is \textit{a priori} fulfilled by choosing the invariants as a measure of deformation, making the material model frame indifferent. This is the deciding factor for constructing the NN as an invariant based model. An added benefit is the guarantee of \textit{material symmetry} since the stress and elasticity tensors are constructed using the derivatives of energy with respect to the symmetric right Cauchy-Green deformation tensor. \textit{Thermodynamic consistency} is fulfilled by constructing the stress tensor from the derivatives of the strain energy, the \textit{normalisation condition} means that in the undeformed case the energy should be zero, e.g. $\psi(\mathbf{F}=\mathbf{I}) = 0$, which is together with \textit{non-negativity of the energy} and \textit{convexity} satisfied through the appropriate choice of activation function and constraining certain weights. \textit{Normalisation of the stress}, i.e. $\mathbf{S}\left(\mathbf{I}\right)=\mathbf{0}$, is satisfied \textit{a priori} by having an incompressible material. The normalisation is fulfilled by the Lagrangian multiplier $p$ which is recovered during training from  Eq.~\eqref{eq:p_definition} (i.e. the plane stress condition that $S_3 = 0$). It should be noted that enforcing the strain energy to be convex and non-decreasing in in the invariants preserves the polyconvexity of the invariants preserving \textit{ellipticity} and ensuring material stability \cite{Klein2023}.

\begin{figure}
	\centering
	\includegraphics{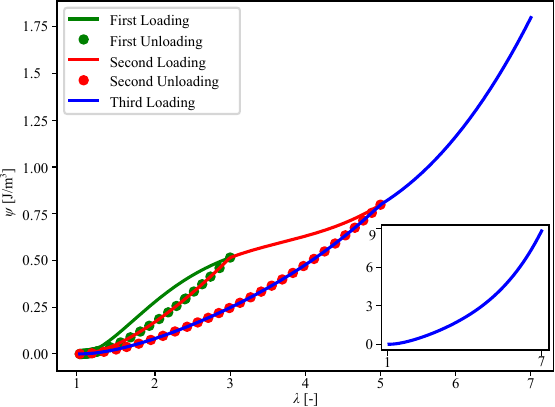}
	\caption{Strain energy curve during loading and unloading for a rubber material with Ogden's law and included Mullins effect. The energy evolution without the Mullins effect, i.e. without damage, is illustrated in the bottom right insert.}
	\label{fig:mullins_energy}
\end{figure}

\textbf{The Mullins effect} is a form of damage that occurs in rubber materials, first observed by \cite{Mullins1948}. When a specimen is unloaded, it follows a different, softer unloading path than when it was first loaded. However, upon reloading the specimen it follows the second path up to the point of unloading where it continues to follow the first loading path, see Fig.~\ref{fig:mullins_energy}. Since damage is a dissipative process the two paths do not coincide, and the area between these two paths can be considered the dissipated energy. There are many models that aim to describe this effect. An overview of the different damage models can be found in \cite{Diani2009}.

The effect can be described by redefining the strain energy by introducing the damage parameter $\zeta$:
\begin{equation}
	\psi(I_1,I_2, \gamma) = \left(1-\zeta(\gamma)\right)\psi_{0}(I_1,I_2),
	\label{eq:mullins_energy}
\end{equation}
where $\psi_{0}$ is the basic (undamaged) strain energy function defining hyperelastic behaviour under consideration such as the Ogden model from Eq.~(\ref{eq:Ogden_model}), or Mooney-Rivlin model from Eq.~(\ref{eq:Mooney-Rivlin}). The damage parameter $\zeta$ is described with the expression
\begin{equation}
	\zeta(\gamma) = \zeta_\infty\left(1-\exp\left(-\frac{\gamma}{\iota}\right)\right),
	\label{eq:zeta_definition}
\end{equation}
where $\gamma$ is the maximum observed value of the undamaged strain energy $\psi_{0}$ during an analysis, $\zeta_\infty$ is the maximum  damage, and $\iota$ is a material dependant constant called the saturation parameter. The damage parameter $\zeta$ can not decrease since it depends on the highest observed value of energy. The stress tensor again follows from Eq.~(\ref{eq:2PK_definition}) as
\begin{equation}
	\mathbf{S} = 2(1-\zeta)\frac{\partial{\psi_{0}}}{\partial{\mathbf{C}}} - p\mathbf{C}^{-1}.
	\label{eq:mullins_stress}
\end{equation}
In Fig.~\ref{fig:mullins_energy} the behaviour is shown for a material with Ogden's law and added Mullins effect with the constants $\zeta_\infty = 0.8$ and $\iota = 1.0$.

\subsection{Preparation of training data}\label{sec:training_prep}

In order to train the neural network, deformation and stress pairs are needed. The $2^\text{nd}$ Piola-Kirchhoff stress is chosen as the stress measure and to characterize the deformation the invariants of the right Cauchy-Green deformation tensor are used. All of the data is generated using the Ogden material model with the constants from Table~\ref{tab:ogden_constants}, $\zeta_\infty = 0.8$ and $\iota = 1$. Three deformation modes in the plane stress condition are used during training, as these are experiments that have been done in order to characterize hyperelastic behaviour \cite{Treloar1944,Ogden1972}:
\begin{itemize}
	\item uniaxial tension
	\begin{equation*}
		\mathbf{F}_\text{uni} =  
		\begin{bmatrix}
			\lambda & 0 & 0 \\
			0 & \lambda^{-0.5} & 0 \\
			0 & 0 & \lambda^{-0.5}
		\end{bmatrix}
	\end{equation*}
	\item equibiaxial tension
	\begin{equation*}
	\mathbf{F}_\text{equi} =  
	\begin{bmatrix}
		\lambda & 0 & 0 \\
		0 & \lambda & 0 \\
		0 & 0 & \lambda^{-2}
	\end{bmatrix}
	\end{equation*}
	\item planar tension
	\begin{equation*}
		\mathbf{F}_\text{PS} =  
		\begin{bmatrix}
			\lambda & 0 & 0 \\
			0 & 1 & 0 \\
			0 & 0 & \lambda^{-1}
		\end{bmatrix}
	\end{equation*}
\end{itemize}

\begin{figure}
	\centering
	\includegraphics{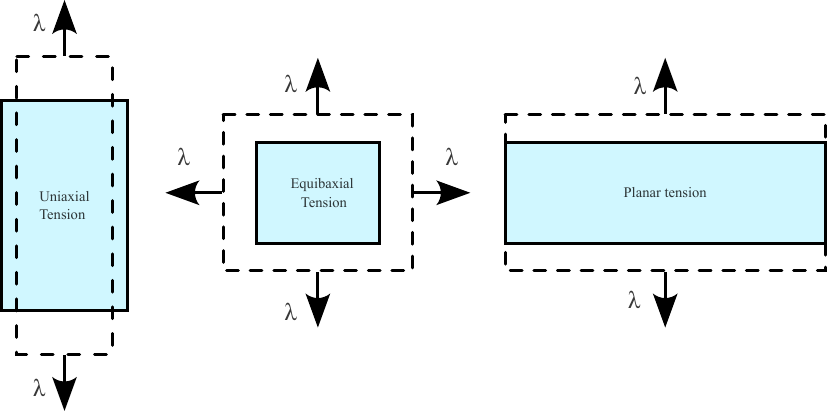}
	\caption{Modes of deformation used during training. The imposed stretch is marked as $\lambda$.}
	\label{fig:deformation_modes}
\end{figure}

In order to fully describe the stress, the Lagrange multiplier $p$ is calculated from the plane stress condition $P_3 = 0$ using Eq.~\eqref{eq:1PK_definition}. Since a historical variable is used when describing the Mullins effect, e.g. $\gamma$ in Eq.~(\ref{eq:zeta_definition}), historical information must also be passed later during training. In Eq.~(\ref{eq:zeta_definition}), it is taken as the maximum observed value of the undamaged strain energy. During training, invariants are used to calculate the historical variable $\gamma$, so the values of invariants for which the maximum strain energy was observed are taken as the historical information to be used later during training. See for example Fig.~\ref{fig:mullins_energy}. During the first unloading phase (or second loading phase) at a stretch of $1\leq\lambda\leq3 $, see Fig.~\ref{fig:mullins_energy}, the historical value of $\lambda$ at which the strain energy $\psi_0$ was maximum is at 3. For that specific loading condition, the current invariants would be calculated using $\lambda$, while for the historically maximum strain energy the invariants will be calculated using $\lambda = 3$. Note the stretches for which $\psi$ and $\psi_0$ are maximum coincide since they are linked by Eq.~\eqref{eq:mullins_energy}.

\begin{figure}
	\centering
	\begin{subfigure}{0.5\textwidth}
	\centering
	\includegraphics[width=\textwidth]{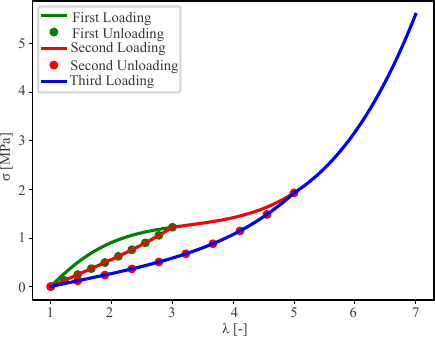}
	\caption{Uniaxial tension.}
	\label{fig:stress_uni}
	\end{subfigure}%
	\begin{subfigure}{0.5\textwidth}
		\centering
		\includegraphics[width=\textwidth]{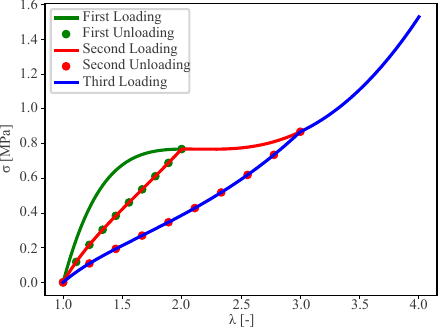}
		\caption{Equibiaxial tension.}
		\label{fig:stress_equi}
	\end{subfigure}
	\begin{subfigure}{0.5\textwidth}
		\centering
		\includegraphics[width=\textwidth]{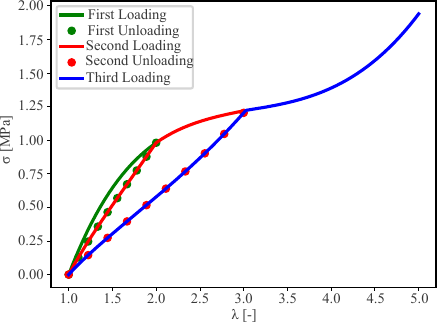}
		\caption{Planar tension.}
		\label{fig:stress_ps}
	\end{subfigure}
	\caption{Stress-stretch plots of the 3 deformation cases, Cauchy stress shown. These are the training data used to train the NNs. 200 data points are taken per each loading/unloading phase.}
	\label{fig:training_data}
\end{figure}

In summary, the input data contains 4 variables $L_s = \left[I_\text{1,max}, I_\text{2,max}, I_1,I_2\right]_s$, and the output data is a set of stress tensors $O_s = \mathbf{S}_s$, with $s \left(= 1,...,N\right)$ being the sample number, and $N$ the total number of samples. Since there are 5 phases (3 loading, 2 unloading) for each deformation mode in Fig.~\ref{fig:training_data}, the total number of data samples is $N = 3000$. This is further split so that $75\%$ is used for actually training the data, and $25\%$ for validation.

As shown in Fig.~\ref{fig:training_data}, the value of the stretch $\lambda$ for each deformation case is:
\begin{itemize}
	\item uniaxial $\lambda$: 1 $\xrightarrow{load} 3 \xrightarrow{unload} 1 \xrightarrow{load} 5 \xrightarrow{unload} 1 \xrightarrow{load} 7$
	\item equibiaxial $\lambda$: 1 $\xrightarrow{load} 2 \xrightarrow{unload} 1 \xrightarrow{load} 3 \xrightarrow{unload} 1 \xrightarrow{load} 4$
	\item planar tension $\lambda$: 1 $\xrightarrow{load} 2 \xrightarrow{unload} 1 \xrightarrow{load} 3 \xrightarrow{unload} 1 \xrightarrow{load} 5$
\end{itemize}
The number of loading/unloading phases can of course be different, as is the case in \cite{Ghaderi2020,Abdusalamov2024}, but here a smaller number was chosen to train the network on less diverse data and demonstrate the ability of the network to capture the underlying behaviour.
\FloatBarrier

\section{Neural network architecture}\label{sec:NN_arch}

\begin{figure}[h]
	\centering
	\includegraphics[scale=1.15]{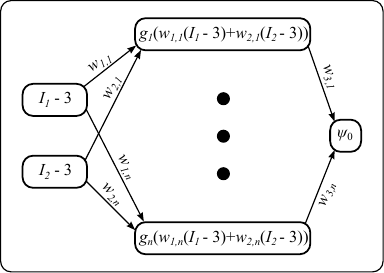}
	\caption{Architecture of the basic FFNN used to model hyperelastic behaviour. This serves as a base for predicting $\psi_{0}$, and is later on used as an "NN block" when explaining the modelling of the Mullins effect with NNs.}
	\label{fig:NN_base_arch}
\end{figure}

Before the Mullins effect can be taken into account, an NN must be constructed that can properly model perfect hyperelastic behaviour, modelled with $\psi_{0}$ in Section~\ref{sec:Mullins_effect}. In this work, a \textit{feed-forward neural network} (FFNN) is used. It is a shallow network consisting of only 1 hidden layer and $n$ neurons within the layer. The illustration of the base NN can be seen in Fig.~\ref{fig:NN_base_arch}. Biases are omitted in order to fulfill the \textit{normalisation condition of energy} and instead of a standard activation function such as the \textit{hyperbolic tangent} or \textit{softplus} function, a function proposed by \cite{Linka2023} was used, namely the \textit{linear exponential} function
\begin{equation}
	g(x) = e^{\alpha x} - 1.
	\label{eq:linear_exponential}
\end{equation}

As pointed out in \cite{Linka2023}, exponential functions are already used when defining hyperelastic behaviour, so the use of an exponential function with a trainable parameter $\alpha$ was tested to obtain an NN that successfully captures hyperelastic behaviour. Unlike \cite{Linka2023}, this work uses a fully connected neural network with mixed products of exponential functions of the invariants $I_1$ and $I_2$. Thus, they are related in this way:

\begin{equation}
	e^{\alpha_i w_{1,i}(I_1-3)}\cdot e^{\alpha_i w_{2,i}(I_2-3)} = e^{\alpha_i \left[ w_{1,i}(I_1-3) + w_{2,i}(I_2-3)\right]},
\end{equation}
so that the invariants are added instead of multiplied. Since the mixed products of $I_1$ and $I_2$ would lead to the occurrence of volumetric components, this approach effectively avoids that problem.

The main advantages of this NN design are its simplicity and the fulfilment of physical constraints mentioned in Section~\ref{sec:Mullins_effect}. \textit{Objectivity} is \textit{a priori} fulfilled by choosing the invariants as a measure of deformation, making the NN frame indifferent. 

\textit{Normalisation condition of the strain energy}, i.e. $\psi(\mathbf{I}) = 0$, is satisfied by setting the inputs as $[I_1-3, I_2-3]$, since in the undeformed case $I_1=I_2=3$. It follows that if $\mathbf{C} = \mathbf{I}$, $g(I_1-3) = e^{0}-1 = 0$. 

\textit{Non-negativity of the strain energy} can only be guaranteed in the case of incompressibility ($\det{\mathbf{F}} = 1$). In this case $I_1 \geq 3, I_2 \geq 3$. With the weights $w_{1,i}, w_{2,i}, \alpha_i$ constrained to be non-negative every activation function $g(x)$ is monotonically increasing and non-negative. If these weights are not restricted then the functions may be non-decreasing and negative (although only up to -1). Additionally, $w_{3,i}$ should be constrained to be non-negative in order to guarantee non-negativity.

\textit{Polyconvexity} is guaranteed by choosing a suitable activation function and ensuring that certain weights are non-negative. Since the output $\psi$ is essentially a sum of all weighted activation functions:
\begin{equation}
	\psi_{0} = \sum_{i=1}^{n}w_{3,i}g_i,
	\label{eq:psi_0_NN}
\end{equation}
only one additional requirement to guarantee the polyconvexity of the NN, which is to make sure that the weights $w_{3,i}$ are enforced to be non-negative. 

\textit{Thermodynamic consistency} is fulfilled by the construction of the NN architecture by choosing the strain energy as an output and the invariants as an input. This respects the definition of stress by deriving the strain energy with respect to an appropriate measure of deformation as given in Eqs.~(\ref{eq:Clausius_Planck_ineq}) and (\ref{eq:2PK_definition}), for further details see the works of \cite{Kalina2021,Linden2023,Linka2023}. 

\textit{Normalisation condition of the stress}, i.e. $\mathbf{S}\left( \mathbf{I} \right) = \mathbf{0}$, is fulfilled by the Lagrangian multiplier $p$ which stems from incompressibility, as discussed in Sec.~\ref{sec:Mullins_effect}. If a compressible material was treated, then an additional normalisation term would need to be added, for further details see \cite{Linden2023}.

The entire process of creating and training the NN was done using the TensorFlow2 library and Python.

\subsection{Accounting for the Mullins effect}\label{sec:NN_Mullins}

\begin{figure}
	\centering
	\includegraphics[width=\textwidth]{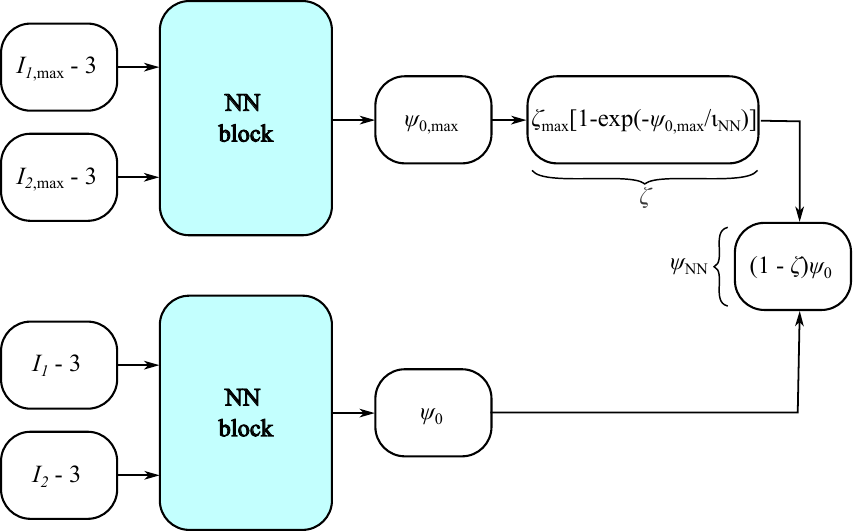}
	\caption{Schematic of the expanded NN that takes into account the Mullins effect. The NN block used here is given in Fig.~\ref{fig:NN_base_arch}. The same block is used twice, so the weights when calculating $\psi_0$ and $\psi_\text{0,max}$ are shared.}
	\label{fig:NN_Mullins_arch}
\end{figure}

Having obtained an NN that can model hyperelastic behaviour, it must now be adapted so that damage is taken into account. This is done by extending the NN from Fig.~\ref{fig:NN_base_arch} to include additional members, in line with how the Mullins effect is included in Section~\ref{sec:Mullins_effect}. Motivated by how conventional hyperelastic models and functions are used to form a \textit{Constitutive Artificial Neural Network} in \cite{Linka2023}, a similar approach has been done here. The calculation of the maximum strain energy and damage variable from Eq.~(\ref{eq:zeta_definition}) was added to the NN and the resulting NN is shown in Fig.~\ref{fig:NN_Mullins_arch}.

The maximum observed value of the undamaged strain energy $\psi_{0}$ can not be measured and trained on. Instead, the entire NN is remodelled so that it predicts the damaged strain energy $\psi$ which can be measured. The maximum undamaged strain energy $\psi_\text{0,max}$ is equivalent to the parameter $\gamma$ from Eq.~(\ref{eq:zeta_definition}), and since it is calculated in the same manner as $\psi_{0}$ the calculations are done in parallel sharing the same weights, hence the repeated use of the same NN block in Fig.~\ref{fig:NN_Mullins_arch}. The trainable parameters in this NN are $w_{1,i}, w_{2,i}$ between the input and hidden layer, $\alpha_i$ in each activation function $g_i$, $w_{3,i}$ between the hidden layer and $\psi_{0}$ and $\psi_\text{0,max}$, and $\zeta_\text{max}$ and $\iota_\text{NN}$ when calculating the damage parameter $\zeta$. The resulting neural network can model hyperelastic behaviour with any level of damage present while fulfilling the constraints in Section~\ref{sec:NN_arch}.

When accounting for the Mullins effect, adding the damage parameter in the form given by Eq.~(\ref{eq:zeta_definition}) does not affect the convexity since the expression $(1-\zeta)$ which can be written as
\begin{equation}
	1-\zeta_\text{max}+\zeta_\text{max}\exp\left(-\frac{\sum_{i=1}^{n}w_{3,i}g_i(I_{1,\text{max}},I_{2,\text{max}})}{\iota_\text{NN}}\right),
	\label{eq:1-zeta_expanded}
\end{equation}
and is always non-negative given the maximum damage constant is $0\leq\zeta_\text{max}\leq1$, and the saturation parameter  is $\iota_\text{NN}>0$.

In the following section, an alternative method for modelling the damage parameter is presented, in which an additional subnetwork is used instead of Eq.~\eqref{eq:1-zeta_expanded}. Please note that throughout the article, the results of the CANN-inspired modelling of the Mullins effect are presented first (Sec.~\ref{sec:training_results},\ref{sec:simple},\ref{sec:solid_disc} and \ref{sec:diabolo}) and the results of the alternative method follow in a corresponding subsection (Secs.~\ref{sec:training_results_subnet},\ref{sec:simple_subnet},\ref{sec:solid_disc_subnet} and \ref{sec:diabolo_subnet}). This is done to ensure clarity given the larger number of results presented.

\subsubsection{Alternative Mullins subnetworks}\label{sec:subnet}

The damage parameter $\zeta$ function is taken from the literature and is one of many ways it can be modelled. However, in a more general case where the shape of the damage evolution can not be assumed it would be advisable to take on a more general approach to modelling the damage parameter. To this end an NN block similar to the one used for the base hyperelastic energy is introduced, which will be referred to as a subnetwork. 

The architecture must be constructed in such a manner to satisfy $\zeta\in [0,1]$ and  $\zeta(\psi_{0,\text{max}} = 0) = 0$. The sigmoid function might seem appropirate, but it fulfills the second requirement only approximately. The hyperbolic tangent however satisfies this condition since tanh(0)=0. The hyperbolic tangent falls within the bounds [-1,1] which is outside of the necessary range of the damage parameter, but for positive values it is also positive and bound at 1. Due to this the tanh satisfies both requirements, although only if a positive argument is passed to it. If the maximum undamaged strain-energy $\psi_{0,\text{max}}$ would be the argument of the function this requirement is satisfied. However, this would not necessarily capture the evolution of the damage parameter. A hidden layer is added between the hyperbolic tangent and the maximum undamaged strain-energy to allow the subnetwork to capture the damage evolution. For the activation function the linear exponential from Eq.~\eqref{eq:linear_exponential} is reused since for $\psi_{0,\text{max}} = 0$ its value is 0 which is passed to the hyperbolic tangent satisfiying  $\zeta(\psi_{0,\text{max}} = 0) = 0$. In the subnetwork the activation functions in the hidden layer are denoted using $h_i$ in order to avoid confusion with the ones used in the NN block in Fig.~\ref{fig:NN_base_arch}. The number of neurons in the hidden layer was set to 5. Finally, in order to further restrict the maximum value of $\zeta$ a trainable parameter $\beta$ is added as a multiplier to the hyperbolic tangent, similar to $\zeta_\infty$ in Eq.~\eqref{eq:zeta_definition}.

\begin{figure}
	\centering
	\begin{subfigure}{0.5\textwidth}
		\centering
		\includegraphics[width=0.95\textwidth]{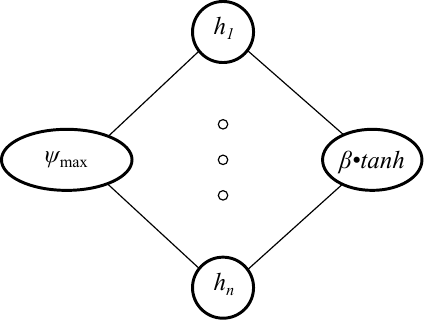}
	\end{subfigure}
	\caption{Alternative subnetwork for modelling the damage parameter $\zeta$ in a more general manner.}
	\label{fig:mullins_subnn_mod_tanh}
\end{figure}

\subsection{Training}\label{sec:NN_training}

Although the NN predicts the strain energy, it is trained exclusively on stresses that could be obtained from standard tests, as mentioned in Section \ref{sec:training_prep}. To this end, a custom loss function must be implemented that calculates the $2^\text{nd}$ Piola-Kirchhoff stress as in Eq.~(\ref{eq:2PK_definition}). This is achieved by using the auto-differentiation tool built into TensorFlow. With Eq.~(\ref{eq:p_definition}),
where the notation $\psi_\text{NN}$ can be used in place of $\psi$ to emphasize that the strain energy is predicted by the NN, the Lagrange multiplier $p$ is derived with respect to the input invariants thus allowing us to calculate stress in plane stress conditions. In case of other conditions such as plane strain or general 3D behaviour, a mixed finite element formulation is used to obtain $p$. Comparing with Fig.~\ref{fig:NN_Mullins_arch}, $\psi_\text{NN}$ is the value of the output neuron $(1-\zeta)\psi_{0}$.

In regards to the convexity condition, it should be noted that although it can be enforced in the entire invariant domain this can lead to a worse approximation than an uncosntrained NN model. This was observed in \cite{Linden2023} where the compressible Neo-Hookean model was used to generate isotropic and transversely isotropic data. The authors noted that the limitation to positive weights (in order to guarantee polyconvexity) reduced the approximation quality of the NN model. As such we relax the polyconvexity criteria so that the weights do not need to be non-negative and train this as one of the NN models. 

The mean squared error between the predicted and "true" stress tensors 
\begin{equation}
	\mathcal{L}_\mathbf{S} = \frac{1}{N}\sum_{i=1}^{N}\left(\mathbf{S}_\text{NN} - \mathbf{S}_\text{true}\right)^2
	\label{eq:loss_function}
\end{equation}
is taken as the measure of loss for the training data. The loss $\mathcal{L}_\mathbf{S}$ does not contain the actual value of energy. This means the NN is trained solely on its derivatives. This is possible due to the normalisation condition since there is a point where the function can be "anchored" at 0 in an undeformed state due to the \textit{normalisation condition of energy} and evolve from it.

Other parameters or settings during training include:
\begin{itemize}
	\item initialisation of the weights $w_{i,j}$ are set using the normal Glorot distribution \cite{Glorot2010}
	\item initialisation of the weights $\alpha_i$ in the activation functions from Eq.~(\ref{eq:linear_exponential}) is set to 0; this was shown to be necessary in order for the NN to train, otherwise the gradients are too large
	\item initialisation of the weights $\zeta_\text{max}$ and $\iota_\text{NN}$ is set to 1 (this is not strictly needed to be 1, it should only be within bounds mentioned below Eq.~(\ref{eq:1-zeta_expanded}) for each of the weights respectively)
	\item the Adam optimizer was used
	\item biases were not included
	\item training was limited to 1 million epochs, but a patience of 20 000 was set to terminate the training if the validation (test) loss does not improve.
\end{itemize}

\subsubsection{Training results}\label{sec:training_results}
\begin{figure}
	\centering
	\includegraphics{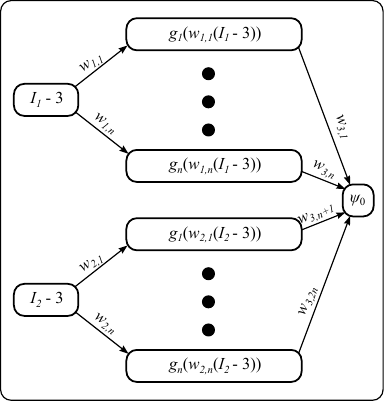}
	\caption{NN block where the invariants are separated rather than added together under the exponential function, case \ref{item_separated}.}
	\label{fig:NN_base_arch_separated}
\end{figure}

In this section, the results of the training are presented, numerical examples and benchmarks are presented in Section~\ref{sec:results}. The influence of the different combinations of training parameters, which can be categorised into 6 different cases, is investigated:
\begin{enumerate}
	\item \textit{polyconvex} NN with the weights $w_{1,i},w_{2,i},w_{3,i}$ and $\alpha_i$ constrained to be non-negative, this is considered a fully constrained NN where all the conditions from Sec.~\ref{sec:NN_arch} are enforced
	\item NN with only $w_{3,i}$ constrained to be non-negative, \textit{polyconvexity} and \textit{non-negativity of the strain energy} are no longer enforced
	\item NN where the weights $w_{3,i}$ are not constrained, while $w_{1,i},w_{2,i},\alpha_i$ are, to see the impact of the constraint on NN training, \textit{polyconvexity} and \textit{non-negativity} are not enforced
	\item NN trained without \textit{polyconvexity} and \textit{non-negativity of the strain energy} being enforced, i.e. none of the weights $w_{1,i},w_{2,i},w_{3,i}$ and $\alpha_i$ are constrained to be non-negative
	\item\label{item_separated} NN where the invariants are not added in the exponent but are passed separately to the hidden layer. This is done to test an NN with the activation function introduced in Eq.~\eqref{eq:linear_exponential} while keeping the strain energy as a sum of individual subfunctions, i.e. $\psi(I_1,I_2) = \psi(I_1) + \psi(I_2)$, see Fig.~\ref{fig:NN_base_arch_separated}.
	\item the isotropic perfectly incompressible CANN model from \cite{Linka2023} was used in place of the NN block presented in Fig.~\ref{fig:NN_base_arch}.
	
\end{enumerate}

In all the cases considered above \textit{objectivity}, \textit{thermodynamic consistency}, and \textit{normalization of the energy} and \textit{stress} are constrained \textit{a priori}.  The polyconvexity of case 1 can be proved by recalling that the composition of twice-differentiable functions $f(\mathbf{x})=f_1(f_2(\mathbf{x}))$ is convex if $f_1$ is convex and non-decreasing in each argument and $f_2$ is convex \cite{Klein2023,boyd2004convex}.  The generalisation of this definition to a function that is the result of many function compositions can be stated as follows. The resulting function is convex if the innermost function is convex, while all other functions are convex and non-decreasing component-wise. The present incompressible context implies that $\psi=\psi\left(\mathbf{F},\text{Cof}\; \mathbf{F}\right)$. Since $I_1$ is convex in $\mathbf{F}$ and $I_2$ is convex in $\text{Cof}\; \mathbf{F})$, then the exponent $\alpha_i \left[w_{1,i}(I_1-3) + w_{2,i}(I_2 - 3)\right]$ must be component-wise convex and non-negative. Alternatively, all the weights ($w_{1,i}$ and $ w_{2,i}$) and trainable parameters $\alpha_i$ must be non-positive. Finally, from Eq.~\ref{eq:psi_0_NN} follows that for the same reason weights $w_{3,i}$ should be non-negative, while the activation function is an exponential function and thus non-negative.

\begin{figure}
	\centering
	\begin{subfigure}{0.5\textwidth}
		\centering
		\includegraphics[width=0.98\textwidth]{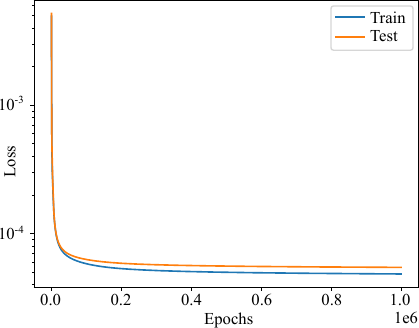}
		\caption{Case 1.}
		\label{fig:fully_constrained_NN_training}
	\end{subfigure}%
	\begin{subfigure}{0.5\textwidth}
		\centering
		\includegraphics[width=0.98\textwidth]{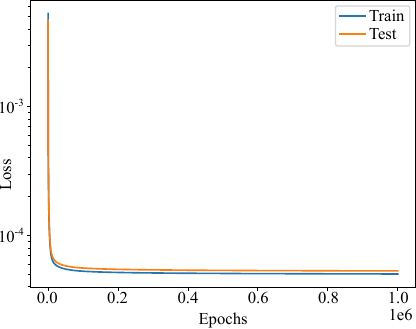}
		\caption{Case 2.}
		\label{fig:convex_NN_training}
	\end{subfigure}
	\begin{subfigure}{0.5\textwidth}
		\centering
		\includegraphics[width=0.98\textwidth]{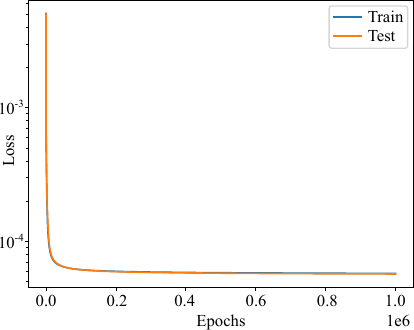}
		\caption{Case 3.}
		\label{fig:last_weigths_free_NN_training}
	\end{subfigure}%
	\begin{subfigure}{0.5\textwidth}
		\centering
		\includegraphics[width=0.98\textwidth]{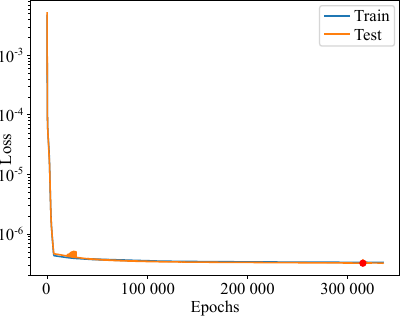}
		\caption{Case 4.}
		\label{fig:nonconvex_NN_training}
	\end{subfigure}
	\begin{subfigure}{0.5\textwidth}
		\centering
		\includegraphics[width=0.98\textwidth]{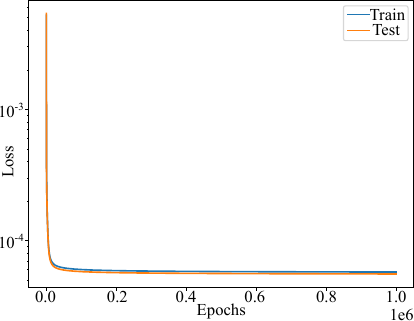}
		\caption{Case 5.}
		\label{fig:separated_NN_training}
	\end{subfigure}%
	\begin{subfigure}{0.5\textwidth}
		\centering
		\includegraphics[width=0.98\textwidth]{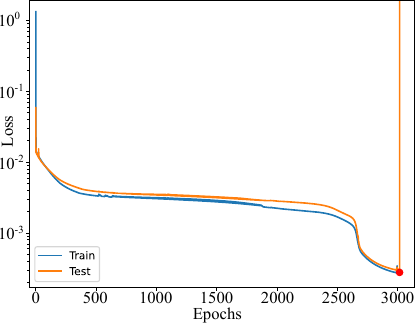}
		\caption{Case 6.}
		\label{fig:CANN_training}
	\end{subfigure}
	\caption{Training results for the different cases of NNs.}
	\label{fig:NN_training_wo_penalty}
\end{figure}

The training curves for allcases are shown in Fig.~\ref{fig:NN_training_wo_penalty}. The influence of the applied constraints to guarantee case 1 and case 5, leads to higher overall losses than other approaches. This means that it does not describe the material behaviour as accurately as the other approaches, but it guarantees polyconvexity on the entire $I_1\text{-}I_2$ domain. Also, the training for cases 1, 2, 3 and 5 lasted until the training limit was reached. It should be noted that the second and third case are not investigated as a viable choice since neither guarantees \textit{polyconvexity} nor \textit{non-negativity of energy} while still introducing constraints. It is investigated to see the influence of various combinations of constraints on the loss. When using the CANN to predict the energy, presented as case 6 in Fig.~\ref{fig:CANN_training}, the NN model underperformed as it could not properly converge during training.

By contrast, in the fourth case, where no constraints or penalties were enforced, the NN achieved a much lower loss with fewer epochs, which is favourable. However, polyconvexity is not guaranteed.

The training results are shown in Fig.~\ref{fig:NN_training_wo_penalty}. In Fig.~\ref{fig:fully_constrained_NN_training}, case 1, the NN is fully constrained, \textit{convexity} and \textit{non-negativity of the strain energy} are guaranteed, the best loss is $5.44\cdot10^{-5}$. Case 2 in Fig.~\ref{fig:convex_NN_training}, where only weights $w_{3,i}$ are constrained to be non-negative has a slightly lower loss of $5.3\cdot10^{-5}$. For the sake of completeness, an NN with the non-negative weights $w_{1,i}, w_{2,i}$ and $\alpha_i$, but unconstrained $w_{3,i}$, case 3, is shown in Fig.~\ref{fig:last_weigths_free_NN_training} so that the influence of the individual constraints can be identified. This NN is not suitable as neither \textit{polyconvexity} nor \textit{non-negativity of the strain energy} are enforced and the loss is even larger at $5.7\cdot10^{-5}$. Removing all constraints leads to the NN obtained in Fig.~\ref{fig:nonconvex_NN_training}, case 4, where the loss is 2 orders of magnitude smaller and is $3.24\cdot10^{-7}$ denoted by the red dot. This is in line with the results obtained in \cite{Linden2023} where the ``unrestricted'' network had the smallest error. It should be noted that in this work \textit{normalisation of energy} and \textit{ stress} are futlfilled \textit{a priori}, while in the work of \cite{Linden2023} a NN without the polyconvexity restriction but with fulfilled normalisation conditions is not presented. In Fig.~\ref{fig:separated_NN_training}, case 5, it is shown that separating the invariants in their own subfunctions instead of adding them as in case 1 through 4 does not impact the training quality since the training loss is still on par with the case 1 at $5.56\cdot10^{-5}$, In Fig.~\ref{fig:CANN_training}, case 6, the NN encountered difficulties during training and reached a loss of $2.77\cdot10^{-4}$ before beginning to diverge. Note that these exact values are not obtained every time a particular case is trained, as the weights $w_{1,i}, w_{2,i}$ and $w_{3,i}$ are randomly initialised during training, but the orders of magnitude remain the same.

Finally,  regarding the \textit{non-negativity of the energy} all the NNs were numerically tested on the invariant domain presented in Fig.~\ref{fig:energy_test_domain} to see if they calculated negative values of energy, and not a single NN has done so. This is attributed to the fulfilment of \textit{thermodynamic consistency} where the NNs were trained on stress data, i.e. the derivatives of energy, thus indirectly enforcing the \textit{non-negativity of energy}.

A comparison of the the results of the different NN cases is presented in Sec.~\ref{sec:simple} where the NN from case 4, i.e. the unconstrained NN, shows the best performance. Because of this, it was used to solve the examples in Sec.~\ref{sec:simple},\ref{sec:solid_disc} and \ref{sec:diabolo}.

\subsubsection{Training results with Mullins subnetworks}\label{sec:training_results_subnet}

Following the results from the previous section the NNs using the subnetworks presented in Sec.~\ref{sec:subnet}, the weights $\alpha_i$ in the activation functions $h_i$ were initialised to 0. Regarding the mulitplier $\beta$, two approaches are presented. In the first one  $\beta$ is fixed to 1 during the training  effectively omitting it as a training variable and removing the restriction of having a maximum damage parameter in the subnetwork akin to $\zeta_\text{max}$, since this would allow the tanh to reach a value of 1. The second allows the parameter $\beta$ to be a trainable parameter that would restrict the damage to a maximum value of less than 1.

\begin{figure}
	\centering
	\begin{subfigure}{0.5\textwidth}
		\centering
		\includegraphics[width=0.98\textwidth]{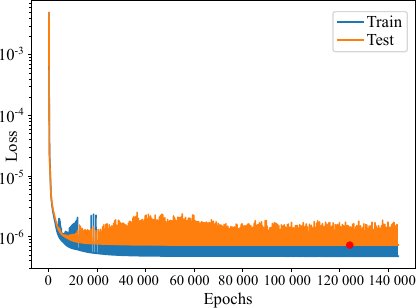}
		\caption{Parameter $\beta$ omitted.}
		\label{fig:training_naive_subNN}
	\end{subfigure}%
	\begin{subfigure}{0.5\textwidth}
		\centering
		\includegraphics[width=0.98\textwidth]{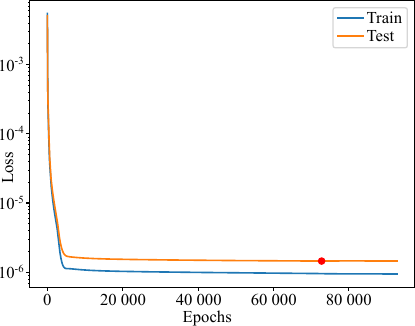}
		\caption{Parameter $\beta$ set as a trainable parameter.}
		\label{fig:training_non_naive_subNN}
	\end{subfigure}
	\caption{Losses during training of the NN with a subnetwork for modelling the damage parameter.}
	\label{fig:subnet_training_results}
\end{figure}

The losses during training are shown in Fig.~\ref{fig:subnet_training_results} where it can be seen that setting $\beta$ as a trainable parameter leads to a smoother training. In fact, the train/test curves in Fig.~\ref{fig:training_naive_subNN} have the most varying losses of all the NNs presented. Also, it takes longer to train the NN compared to Fig.~\ref{fig:training_non_naive_subNN}. However, the upshot is that the loss is smaller at $7.24\cdot10^{-7}$ compared to $1.44\cdot10^{-6}$ when $\beta$ is set as a trainable parameter. Additionally, after training the parameter $\beta$ was obtained as 0.86, although this may vary slightly each training run.

\section{Numerical examples}\label{sec:results}

To evaluate performance of the proposed framework, three  benchmarks are considered. A first verification of correctness is done by simulating uniaxial, equibiaxial, and planar tension tests in Abaqus. The second numerical example is a simulation of a rolling solid rubber disc. This example comes from Abaqus' example benchmarks and uses the Mullins effect \cite{Abaqus614}. The third numerical example comes from \cite{Chagnon2006} and represents a body that is loaded axially and then rotated along the longitudinal axis. It should be noted that all examples use the Ogden model with Mullins effect as a reference. Although the NN was trained only on the 3 basic homogenoues deformation modes in the plane stress condition, all the numerical examples are performed in a 3D setting where the Lagrange multiplier $p$ can not be obtained as simply as before from a the plane stress boundary condition, but is instead obtained numerically. In addition, deformation modes that occur in the second and third examples are no longer simple homogenous deformations.

All numerical examples are performed with a fully incompressible hybrid formulation in Abaqus, the hexahedral element type C3D8H is used. The NN model is implemented in Abaqus using the subroutine UHYPER. The explicit expressions for the derivatives of the activation functions with respect to the inputs of the NN are given in \ref{sec:implementation_details}. In this way, the subroutine UHYPER can be coded so that it directly returns the values of all necessary derivatives. For details on the implementation see \ref{sec:implementation_details}.

\subsection{Simple verification tests}\label{sec:simple}

A first step in testing the NN material model is to simulate the aforementioned tests in Section~\ref{sec:training_prep} with finite element simulations. Different load parameters are used to avoid simply reproducing the training data. To this end the following stretches are prescribed:
\begin{itemize}
	\item uniaxial $\lambda: 1 \xrightarrow{load} 2 \xrightarrow{unload} 1 \xrightarrow{load} 3 \xrightarrow{unload} 1 \xrightarrow{load} 6$
	\item equibiaxial $\lambda: 1 \xrightarrow{load} 1.5 \xrightarrow{unload} 1 \xrightarrow{load} 2.5 \xrightarrow{unload} 1 \xrightarrow{load} 4 $
	\item  planar tension $ \lambda: 1 \xrightarrow{load} 1.75 \xrightarrow{unload} 1 \xrightarrow{load} 2.5 \xrightarrow{unload} 1 \xrightarrow{load} 5$
\end{itemize}

\begin{figure}[h!]
	\centering
	\begin{subfigure}{0.5\textwidth}
		\centering
		\includegraphics[width=\textwidth]{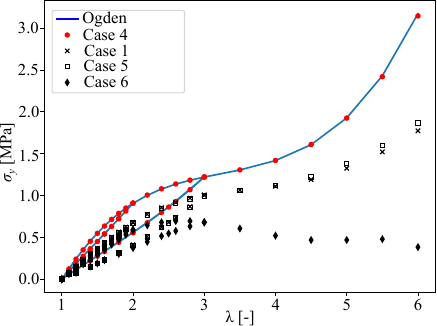}
		\caption{Uniaxial test.}
		\label{fig:stress_abaqus_uni}
	\end{subfigure}%
	\begin{subfigure}{0.5\textwidth}
		\centering
		\includegraphics[width=0.97\textwidth]{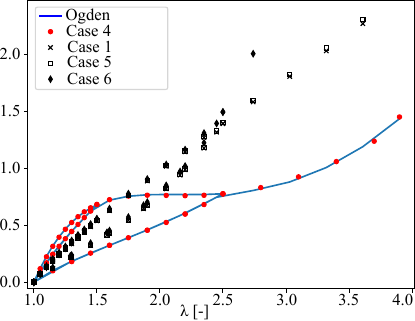}
		\caption{Equibiaxial test.}
		\label{fig:stress_abaqus_equi}
	\end{subfigure}
	\begin{subfigure}{0.5\textwidth}
		\centering
		\includegraphics{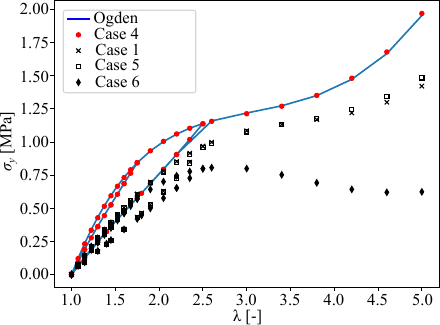}
		\caption{Planar tension test.}
		\label{fig:stress_abaqus_ps}
	\end{subfigure}
	\caption{Comparison of stress-stretch plots of the 3 deformation cases with the results from Abaqus.}
	\label{fig:abaqus_simple_tests}
\end{figure}

\begin{figure}
	\centering
	\begin{subfigure}{0.5\textwidth}
		\centering
		\includegraphics[width=\textwidth]{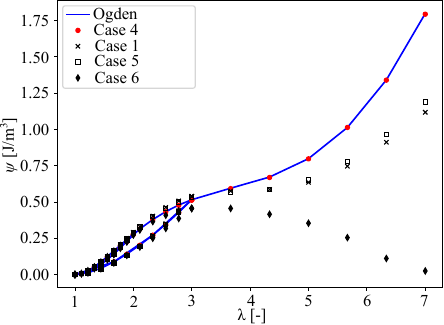}
		\caption{Mullins effect included.}
		\label{fig:uni_mullins_energy}
	\end{subfigure}%
	\begin{subfigure}{0.5\textwidth}
		\centering
		\includegraphics[width=0.98\textwidth]{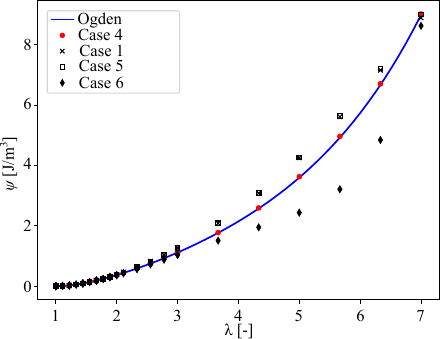}
		\caption{Undamaged strain energy $\psi_{0}$.}
		\label{fig:uni_undamaged_energy}
	\end{subfigure}
	\begin{subfigure}{0.5\textwidth}
		\centering
		\includegraphics{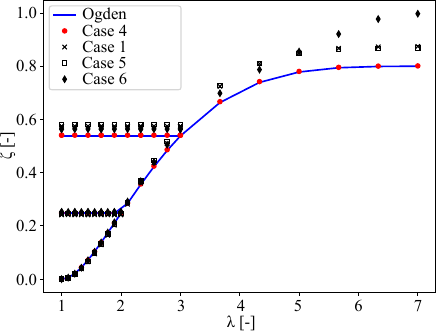}
		\caption{Evolution of the damage parameter $\zeta$ for uniaxial tension.}
		\label{fig:uni_damage_evolution}
	\end{subfigure}
	\caption{Plots of the strain energy, both with and without damage, and the damage parameter evolution for uniaxial tension. These plots demonstrate the ability of the different NNs to predict the evolution of the strain energy without being trained on it, as well as the evolution of $\psi_0$ and $\zeta$.}
	\label{fig:psi0_and_zeta}
\end{figure}

For all 3 examples, a cube with an edge of 100 mm is used, with 5 hexahedral FEs per edge. The results in Fig.~\ref{fig:abaqus_simple_tests} show the Cauchy stress in the \textit{y} direction versus the prescribed stretch. The median relative errors for the uniaxial, equibiaxial and planar tension tests are $0.35\%$, $0.89\%$ and $0.34\%$ respectively. The NN model from case 4 is able to capture the behaviour well on different stretch levels than it was trained on. It passes the simple stress-stretch verification tests, however it might be interesting to see whether the NN reproduces the same material behaviour it was trained on. As for cases 1, 5 and 6, discrepancy from the Ogden model is clearly visible for the complete range of stretches $\lambda$. In addition to the stress-stretch plots, it is of interest to see if the NN successfully captured the strain energy function $\psi$ and the undamaged strain energy $\psi_0$, as well as the evolution of the damage variable $\zeta$. These results are given in Fig.~\ref{fig:psi0_and_zeta}. Again, the NN from case 4 is the best performing one, while the remaining models exhibit reasonable accruracy only for lower stretches ($\lambda<3$).

\begin{figure}
	\centering
	\includegraphics{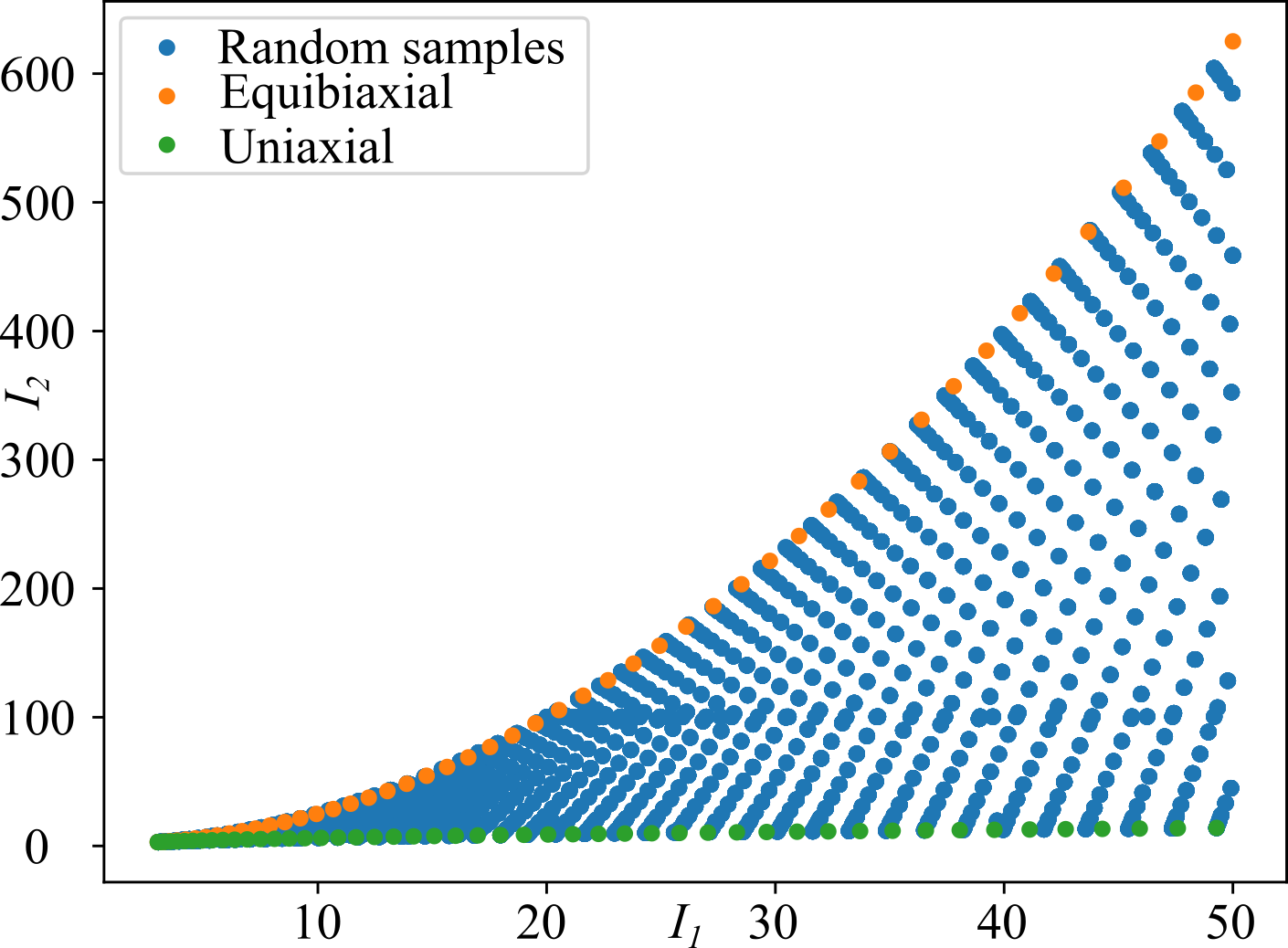}
	\caption{View of the invariant domain on which the energy is tested. Random samples are gathered by sampling deformation gradients, uniaxial and equibiaxial curves are overlayed.}
	\label{fig:energy_test_domain}
\end{figure}

\begin{figure}
	\centering
	\includegraphics{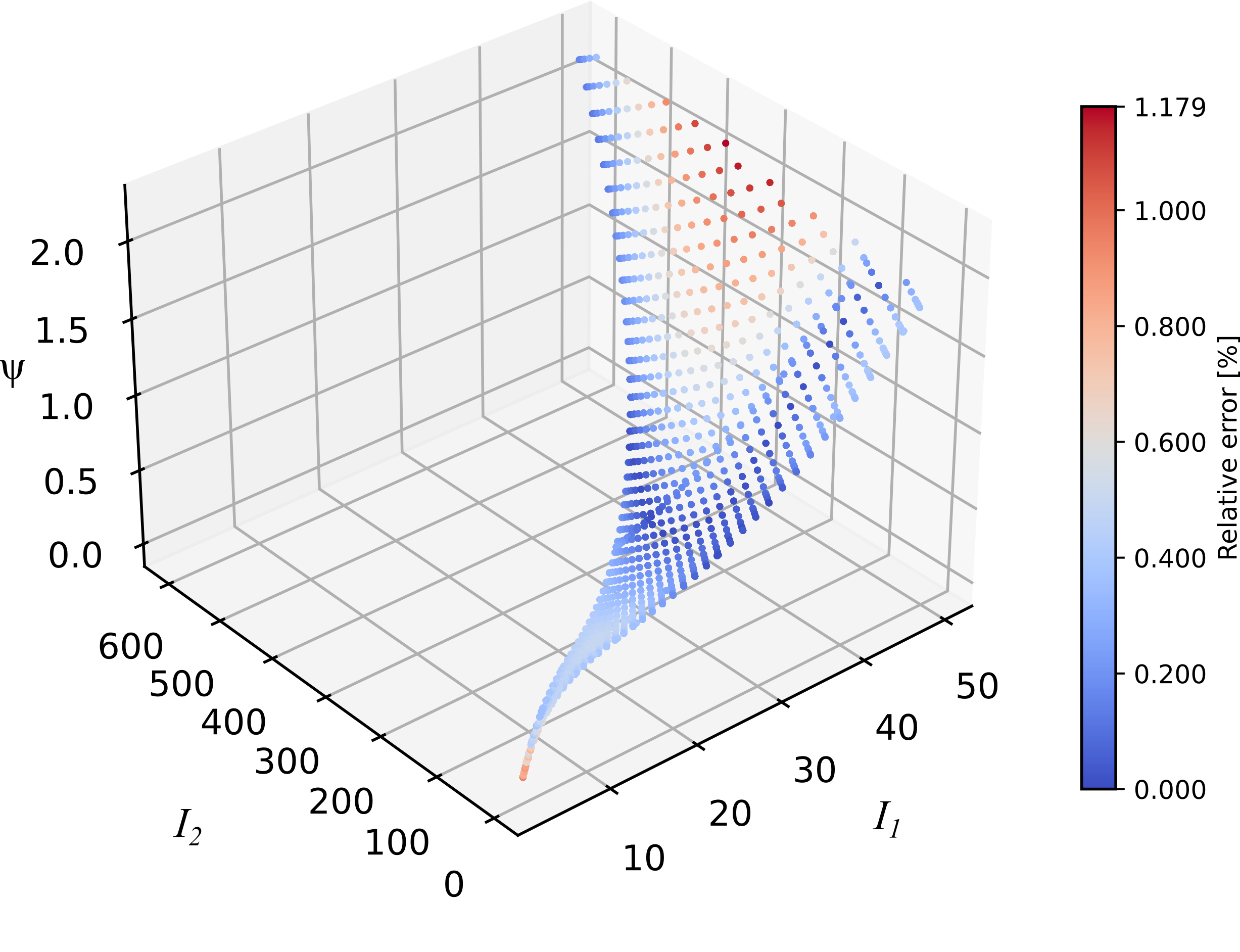}
	\caption{Relative error of the predicted Mullins energy over the invariant domain. Results of the NN with specialised damage function. The points are coloured according to the relative error of the NN with respect to the Ogden model.}
	\label{fig:Error_over_invariant_domain_NN}
\end{figure}

In order to evaluate how well the NN captured the energy further away from the bounds of the domain on which it was trained on a 3D plot over the $I_1$-$I_2$ domain is shown in Fig.~\ref{fig:Error_over_invariant_domain_NN}. The maximum error of 1.18\% is located the furthest from the bounds. The domain, see Fig.~\ref{fig:energy_test_domain}, was create by randomly sampling deformation gradients where the 2 principal stretches $\lambda_1$ and $\lambda_2$ were randomly assigned values from 0.1 to 7, with the condition that the invariants must not exceed the bounds of the uniaxial and equibiaxial deformations (i.e. $I_1\leq50$ and $I_2\leq625$).

\subsubsection{Simple tests - Mullins subnetwork}\label{sec:simple_subnet}
In this section, the results of using the subnetworks from Sec.~\ref{sec:subnet} for uniaxial tension to investigate how well the subnetworks capture the underlying damage. The NN from case 4 is used to predict the underlying behaviour of $\psi_0$. A comparison of the two approaches is shown in Fig.~\ref{fig:subnet_comparison}. Both approaches correctly predict the stresses up to $\lambda=7$, with errors of $1.6\%$ and $4.3\%$ in Fig~\ref{fig:naive_uniaxial_stress} and Fig.~\ref{fig:non_naive_uniaxial_stress} respectively. However, a look at the other figures shows that it is disadvantageous to make the parameter $\beta$ trainable, as it does not correctly capture the underlying behaviour in the entire range of interest, but only up to a value of about $\lambda=5$. After that, it starts to predict the Mullins energy and the undamaged energy as well as the damage variable poorly. The energy in Fig.~\ref{fig:non_naive_uniaxial_mullins} is underestimated, while the undamaged energy in Fig.~\ref{fig:non_naive_uniaxial_undamaged} is overestimated, as is the damage variable $\zeta$ in Fig.~\ref{fig:non_naive_uniaxial_damage}. The value obtained for the parameter $\beta$ is 0.86 and is therefore above the maximum damage constant of $\zeta_\infty = 0.8$ with which the data was generated. For a stretch of $\lambda=7$ in Fig.~\ref{fig:non_naive_uniaxial_damage}, the damage variable reaches the maximum value of 0.86 and thus the NN recovers a similar behaviour to the original one by correctly replicating the stresses but not the damage and energy.
On the other hand, if the parameter $\beta$ is omitted from the training, the NN captures exactly the underlying behaviour that was used to generate the training data. In Figs.~\ref{fig:naive_uniaxial_mullins},\ref{fig:naive_uniaxial_undamaged} and \ref{fig:naive_uniaxial_damage}, the Mullins energy and undeformed energy are correctly captured, as is the damage variable with the respective errors at $\lambda=7$ of $0.5\%,1.44\%$ and $0.2\%$.
\begin{figure}
	\centering
	\begin{subfigure}{0.5\textwidth}
		\centering
		\includegraphics[width=\textwidth]{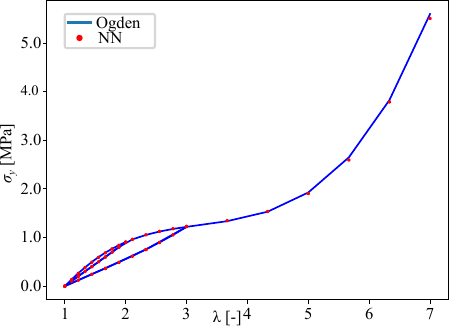}
		\caption{Subnetwork uniaxial Cauchy stress, $\beta=1$.}
		\label{fig:naive_uniaxial_stress}
	\end{subfigure}%
	\begin{subfigure}{0.5\textwidth}
		\centering
		\includegraphics[width=0.945\textwidth]{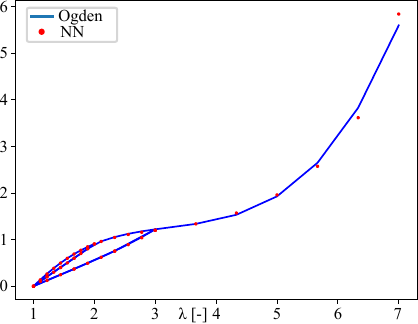}
		\caption{Subnetwork uniaxial Cauchy stress, $\beta$ is trainable.}
		\label{fig:non_naive_uniaxial_stress}
	\end{subfigure}
	\begin{subfigure}{0.5\textwidth}
		\centering
		\includegraphics[width=\textwidth]{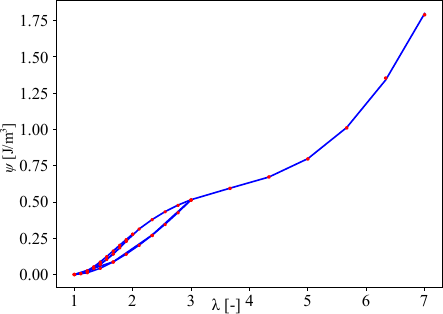}
		\caption{Subnetwork Mullins energy, $\beta=1$.}
		\label{fig:naive_uniaxial_mullins}
	\end{subfigure}%
	\begin{subfigure}{0.5\textwidth}
		\centering
		\includegraphics[width=\textwidth]{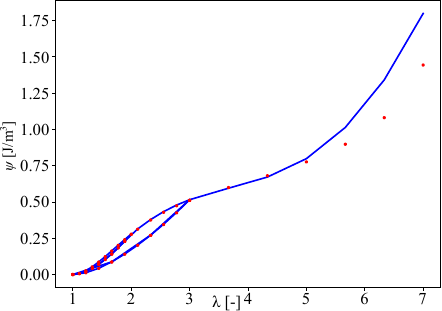}
		\caption{Subnetwork Mullins energy, $\beta$ is trainable.}
		\label{fig:non_naive_uniaxial_mullins}
	\end{subfigure}
	\begin{subfigure}{0.5\textwidth}
		\centering
		\includegraphics[width=\textwidth]{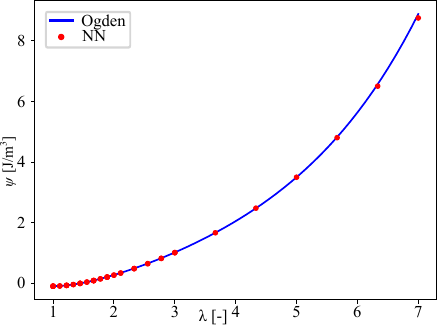}
		\caption{Subnetwork undamaged energy, $\beta=1$.}
		\label{fig:naive_uniaxial_undamaged}
	\end{subfigure}%
	\begin{subfigure}{0.5\textwidth}
		\centering
		\includegraphics[width=\textwidth]{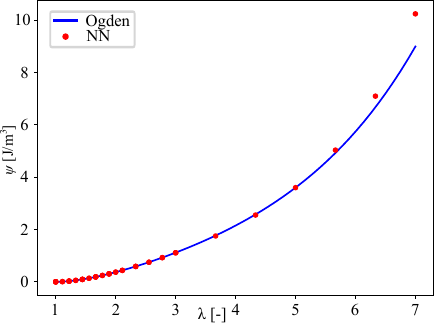}
		\caption{Subnetwork undamaged energy, $\beta$ is trainable.}
		\label{fig:non_naive_uniaxial_undamaged}
	\end{subfigure}
\end{figure}
\begin{figure}\ContinuedFloat
	\begin{subfigure}{0.5\textwidth}
		\centering
		\includegraphics[width=0.99\textwidth]{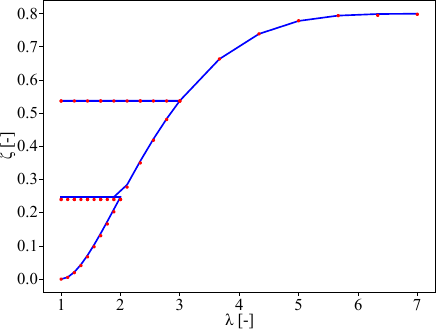}
		\caption{Subnetwork damage evolution, $\beta=1$.}
		\label{fig:naive_uniaxial_damage}
	\end{subfigure}%
	\begin{subfigure}{0.5\textwidth}
		\centering
		\includegraphics[width=0.99\textwidth]{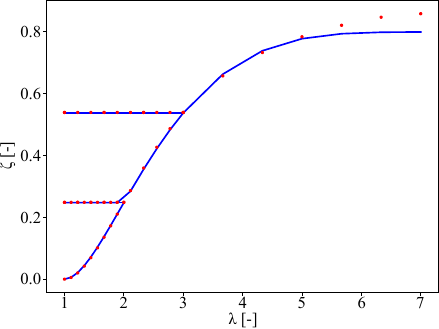}
		\caption{Subnetwork damage evolution, $\beta$ is trainable.}
		\label{fig:non_naive_uniaxial_damage}
	\end{subfigure}
	\caption{Results of Cauchy stress, damaged and undamaged energy, and damage parameter evolution for an NN with a subnetwork. Figures on the left represent the approach where $\beta$ is not a trainable parameter while the figures on the right represent the approach where $\beta$ is a trainable parameter, with a value of $\beta = 0.86$.}
	\label{fig:subnet_comparison}
\end{figure}

\begin{figure}
	\centering
	\includegraphics{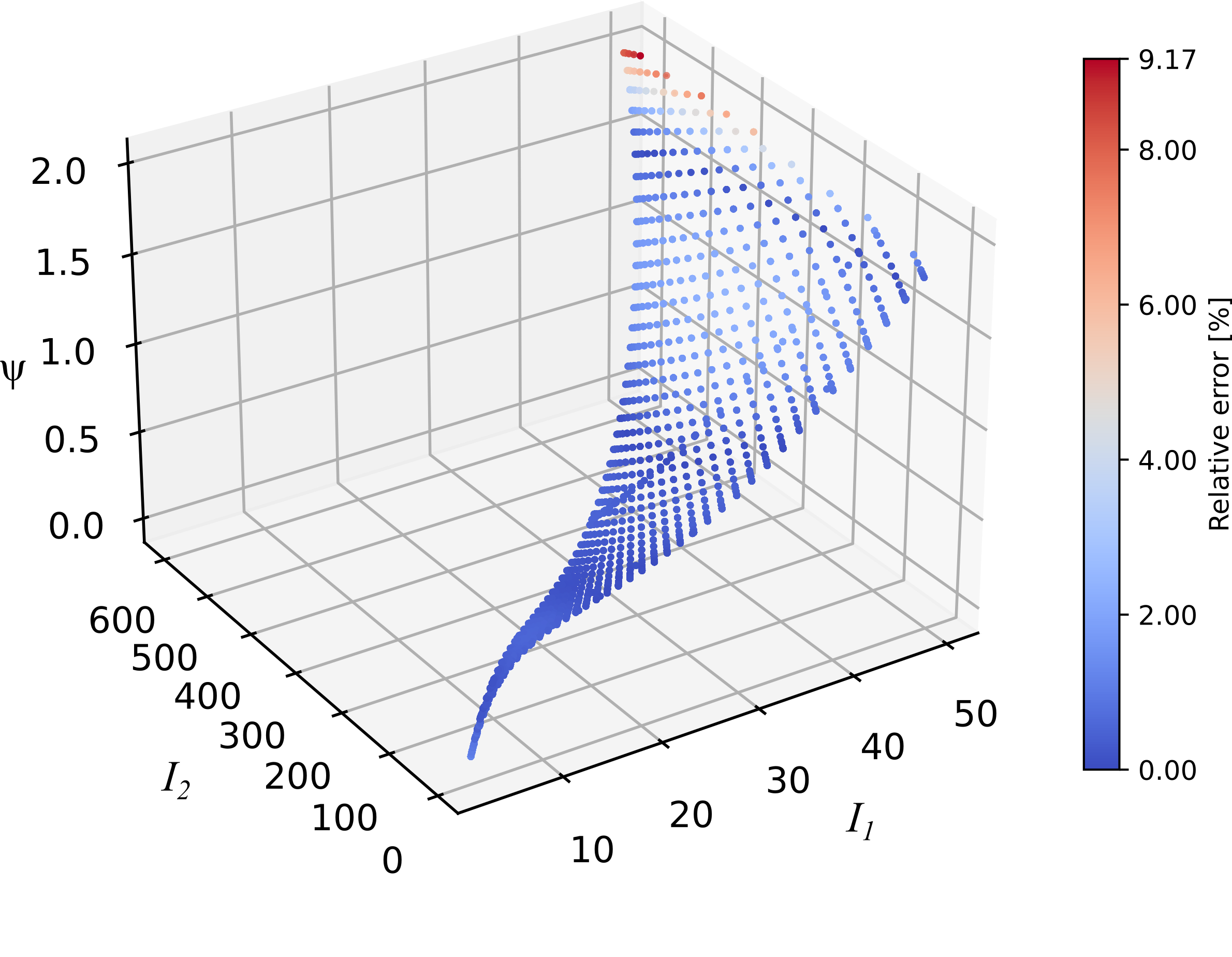}
	\caption{Relative error of the predicted Mullins energy over the invariant domain. Results are shown for the NN with a subnetwork and the parameter $\beta$ omitted. The points are coloured according to the relative error of the NN with respect to the Ogden model.}
	\label{fig:Error_over_invariant_domain_naive_subNN}	
\end{figure}

\begin{figure}
	\centering
	\includegraphics{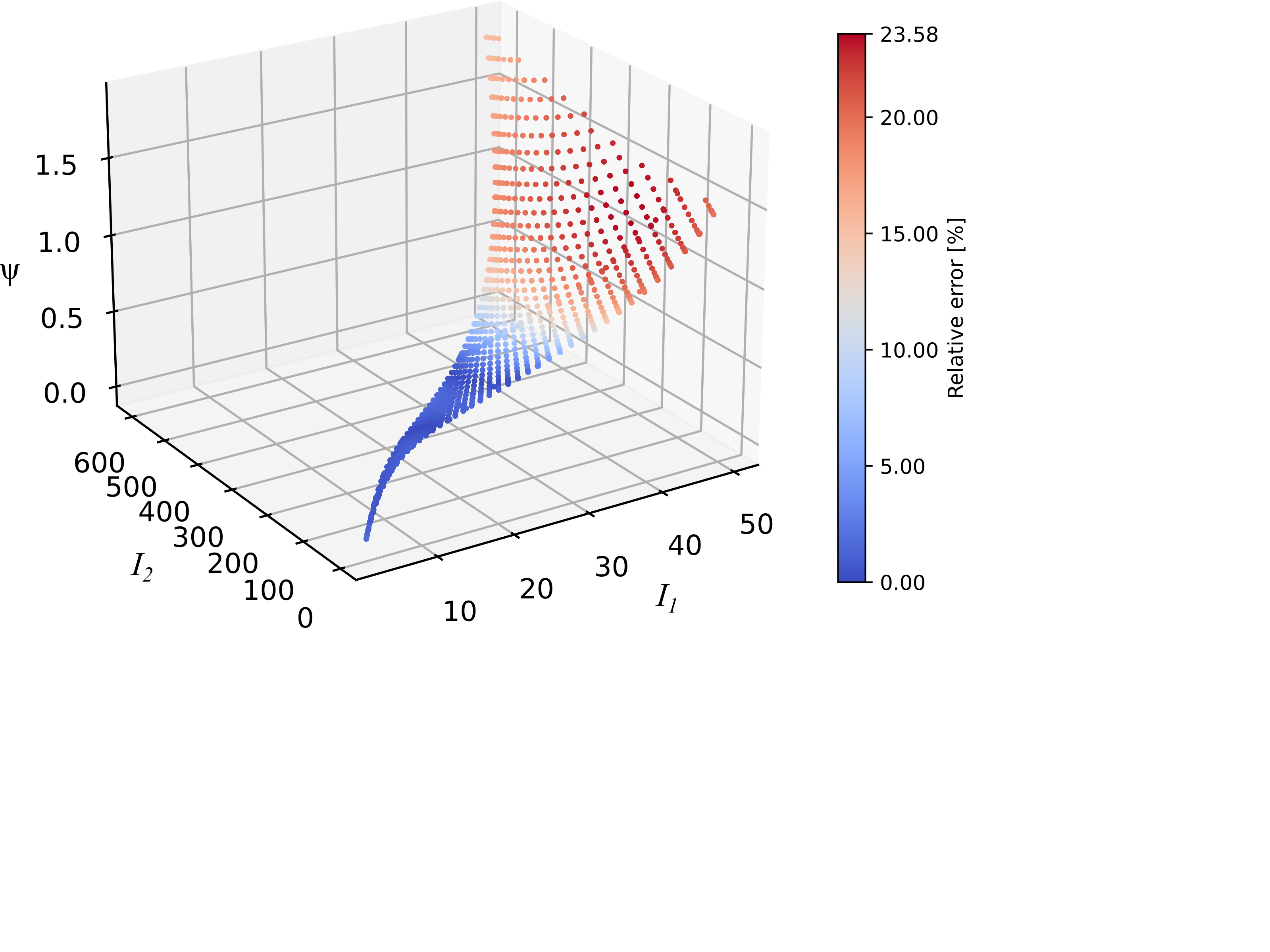}
	\caption{Relative error of the predicted Mullins energy over the invariant domain. Results are shown for the NN with a subnetwork and the parameter $\beta$ included ($\beta = 0.86$). The points are coloured according to the relative error of the NN with respect to the Ogden model.}
	\label{fig:Error_over_invariant_domain_non_naive_subNN}	
\end{figure}

Additionally, in the same manner as in Fig.~\ref{fig:Error_over_invariant_domain_NN} the relative errors of the strain energy for the NNs with a subnetwork are presented in Figs.~\ref{fig:Error_over_invariant_domain_naive_subNN}~and~\ref{fig:Error_over_invariant_domain_non_naive_subNN}.


\subsubsection{Cyclic testing}
\begin{figure}
	\centering
	\includegraphics{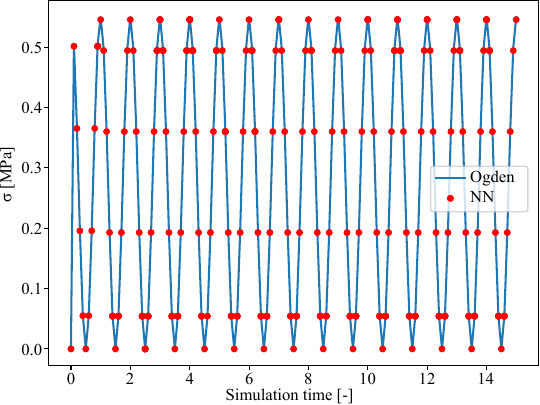}
	\caption{Cyclic test with a constant amplitude, first 15 cycles shown for illustration, the test was performed for 1000 cycles.}
	\label{fig:cyclic_testing_one}
\end{figure}

\begin{figure}
	\centering
	\includegraphics{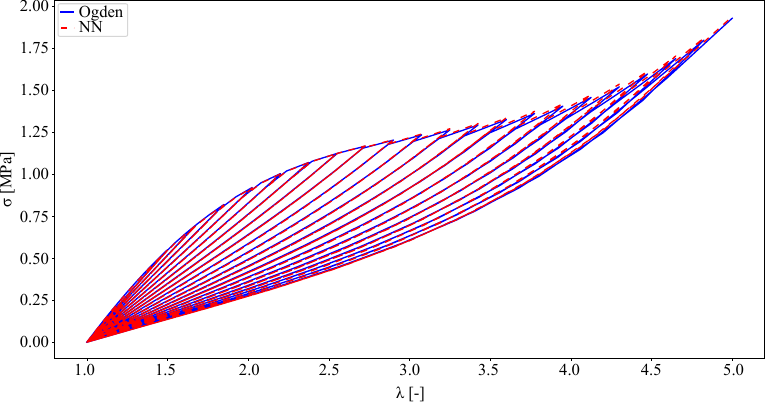}
	\caption{Cyclic test with progressively increasing loads, 20 loading cycles performed.}
	\label{fig:cyclic_testing_two}
\end{figure}

Another simple test to see if the NN has learned the underlying behaviour correctly would be a cyclic loading test. Two cyclic loading tests were performed, one with a constant sinusoidal amplitude and the second in which the loading cycles became progressively larger to demonstrate the Mullins effect. The first test was used to check that the NN always gave correct results so that no errors occurred, while the second test was used to see how well the Mullins effect was captured. The results shown refer to the unconstrained NN (case 4 from Sec.~\ref{sec:training_results}), but they also apply to the NNs with a subnetwork. The first cyclic test was performed for 1000 cycles and ran without problems. The results for the first 15 cycles are shown in Fig.~\ref{fig:cyclic_testing_one}. The second test was performed for 20 loading cycles, with small increments in the final stretch each time. The Mullins effect was successfully captured, as can be seen in Fig.~\ref{fig:cyclic_testing_two}.


\subsection{Solid rubber disc}\label{sec:solid_disc}

\begin{figure}[h!]
	\centering
	\includegraphics{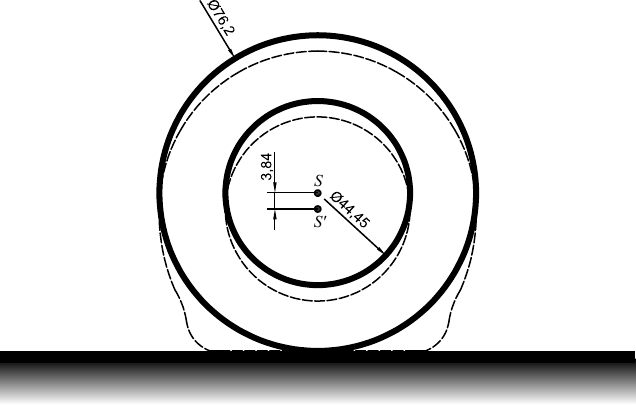}
	\caption{Solid rubber disc example from Abaqus. This is a simplified model of a vehicle tyre. The dimensions are given in millimetres. The dashed line represents the approximate deformation after the original geometry (solid lines) has been pressed into the ground by 3.84 mm. The thickness of the solid rubber disc is 17.78 mm. The centre of the disc marked with $S$ is the initial position at which the boundary condition displacement/rotation was prescribed, and $S'$ denotes the position after pressing in.}
	\label{fig:solid_disc_dimensions}
\end{figure}

The example of the solid disc comes from Abaqus \cite{Abaqus614}. Its dimensions are given in Fig.~\ref{fig:solid_disc_dimensions}. A rigid coupling is created in the centre of the disc, on which a downward displacement of 3.84 mm and a rotation of one full circle are prescribed. It is pressed against an analytical rigid surface with frictionless contact. The aim of this example is to test the NN model against loading conditions and deformation modes it has not been trained on and test whether it has learned full hyperelastic behaviour from previously shown simple examples.

\begin{figure}[h!]
	\centering
	\includegraphics{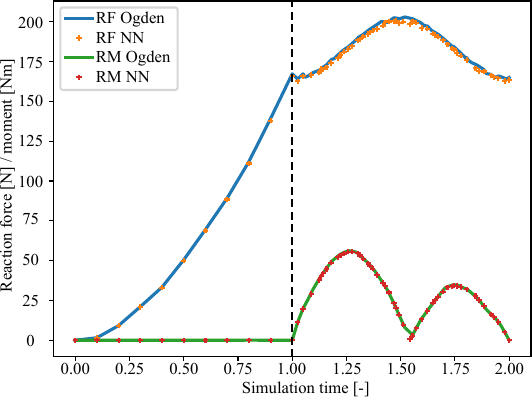}
	\caption{Reaction forces and moments at the central reference point at which the displacement/rotation was prescribed. The vertical dashed line marks the end of the displacement step and the beginning of the rotation.}
	\label{fig:solid_disc_reactions}
\end{figure}

The reaction forces of the Ogden reference model compared to the NN model are shown in Fig.~\ref{fig:solid_disc_reactions}. They are measured at the central reference point where the displacement and rotation are prescribed. It shows excellent agreement with the median relative error for the reaction forces of $0.92\%$ and for the reaction moments of $1.03\%$.

\begin{figure}
	\centering
	\begin{subfigure}{\textwidth}
		\centering
		\includegraphics{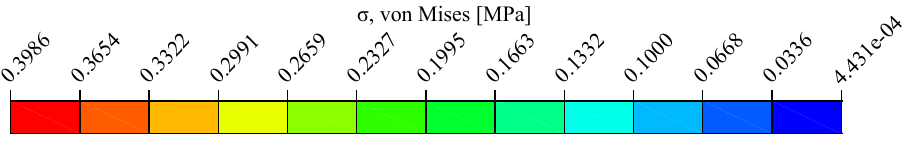}
	\end{subfigure}
	\begin{subfigure}{0.5\textwidth}
		\centering
		\includegraphics[width=0.95\textwidth]{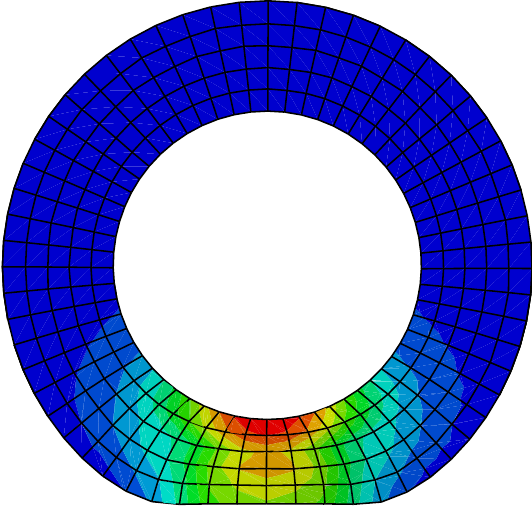}
		\caption{NN solution, max 0.3962 MPa, min 4.431e-4 MPa.}
		\label{fig:solid_disc_mises_NN}
	\end{subfigure}%
	\begin{subfigure}{0.5\textwidth}
		\centering
		\includegraphics[width=0.95\textwidth]{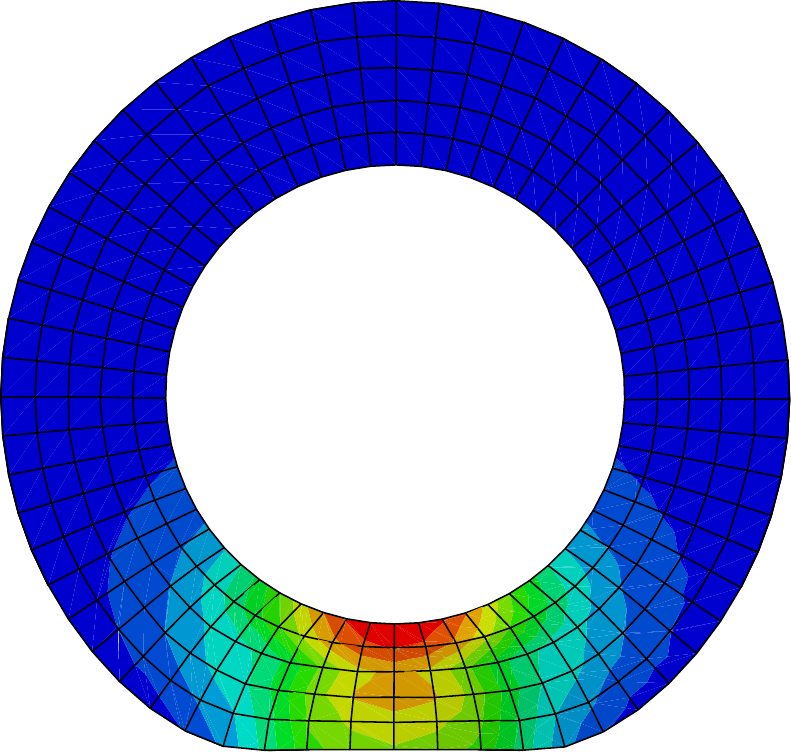}
		\caption{Ogden solution, max 0.3986 MPa, min 4.47e-4 MPa.}
		\label{fig:solid_disc_mises_Ogden}
	\end{subfigure}
	\begin{subfigure}{0.5\textwidth}
		\centering
		\includegraphics[width=\textwidth]{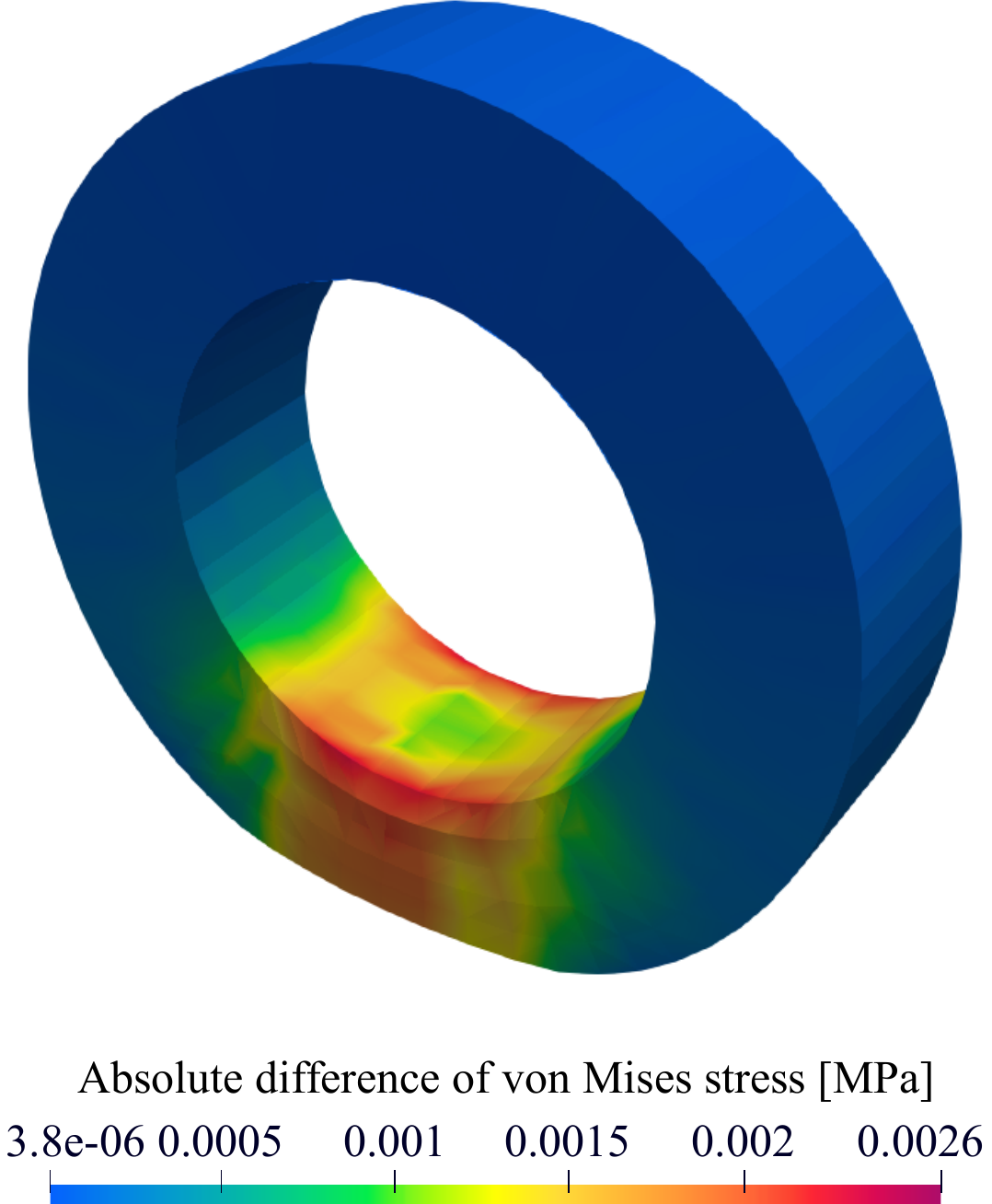}
		\caption{}
		\label{fig:absolute_stress_difference}
	\end{subfigure}
	\caption{Von Mises stress plots of the NN and referent solution, the absolute difference of the solutions is given as well. The relative error of maximum stress is 0.6\% and of the minimum 0.88\%. }
	\label{fig:solid_disc_mises}
\end{figure}

In addition to the reaction force and moment diagram, the von Mises stress plots are shown in Fig.~\ref{fig:solid_disc_mises}. The plots are in agreement and the relative error of the maximum stress is about 0.6\%.

\begin{figure}
	\centering
	\begin{subfigure}{0.5\textwidth}
		\centering
		\includegraphics[width=0.95\textwidth]{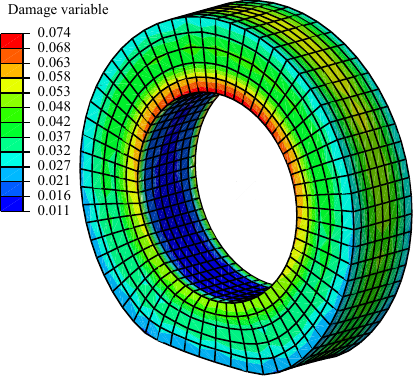}
		\caption{NN model.}
		\label{fig:damage_nn_solution}
	\end{subfigure}%
	\begin{subfigure}{0.5\textwidth}
		\centering
		\includegraphics[width=0.95\textwidth]{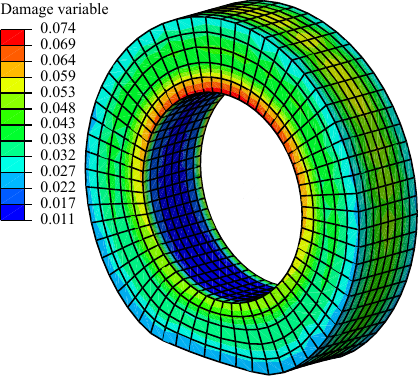}
		\caption{Ogden model.}
		\label{fig:damage_ogden_solution}
	\end{subfigure}
	\caption{Plots of damage variable $\zeta$ of the NN model and reference Ogden model.}
	\label{fig:damage_comparison_solid_disc}
\end{figure}

Apart from the current stress values, the damage at a certain part of a body might be a quantity of interest. For comparison, the plots of the damage variables across the solid disc are given in Fig.~\ref{fig:damage_comparison_solid_disc}. Comparing the damage plots it can be seen that the values of the damage variable $\zeta$ from Eqs.~(\ref{eq:mullins_energy})~and~(\ref{eq:zeta_definition}) are in agreement. They are slightly differently spread out  since the underlying material behaviour is not the same and the values of $\zeta_\infty$ and $\iota$ learned by the NN are not identical. This difference is, however, barely noticable and the NN model has succesfully captured the evolution of the damage, as it has on the simpler example in Fig.~\ref{fig:uni_damage_evolution}.

In summary, the presented results for this case where more complex loading and deformation are present have shown that the NN model captures the Mullins effect and hyperelastic behaviour.

\subsubsection{Solid rubber disc - Mullins subnetwork}\label{sec:solid_disc_subnet}

\begin{figure}
	\centering
	\begin{subfigure}{0.5\textwidth}
		\centering
		\includegraphics[width=\textwidth]{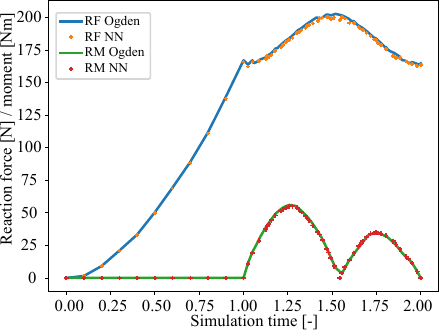}
		\caption{NN with subnetwork, $\beta$ omitted.}
		\label{fig:solid_disc_naive_reactions}
	\end{subfigure}%
	\begin{subfigure}{0.5\textwidth}
		\centering
		\includegraphics[width=0.97\textwidth]{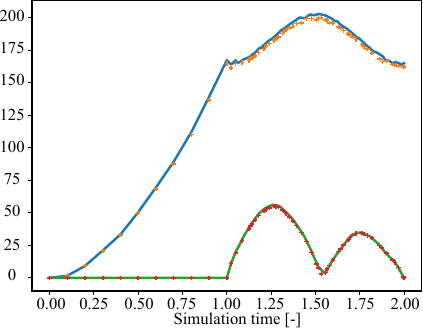}
		\caption{NN with subnetwork, $\beta$ is trainable.}
		\label{fig:solid_disc_non_naive_reactions}
	\end{subfigure}
	\caption{Reaction forces and moments at the central reference point at which the displacement/rotation was prescribed. Results are shown for the NNs with a subnetwork that models the damage. The vertical dashed line marks the end of the displacement step and the beginning of the rotation.}
	\label{fig:solid_disc_subnet_reactions}
\end{figure}

In this section the results of the subnetworks are shortly presented. Both subnetwork approaches correctly capture the behaviour with the reaction forces and moments being shown in Fig.~\ref{fig:solid_disc_subnet_reactions}. The median errors of the reaction force and moment for the subnetwork without $\beta$ are 0.9\% and 1.8\%, and for the subnetwork with $\beta$ the median errors are 1.5\% and 1.74\%. The errors are very close to the ones for the NN with a specialised $\zeta$ function in Fig.~\ref{fig:solid_disc_reactions}.
The stress and damage variable plots are not presented since the results are in line with what has been already shown in the previous section in Figs.~\ref{fig:solid_disc_mises}~and~\ref{fig:damage_comparison_solid_disc}, following the trend of the reaction force and moment.

\subsection{Diabolo}\label{sec:diabolo}

\begin{figure}[h!]
	\centering
	\includegraphics{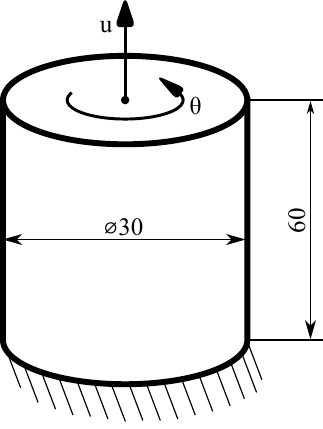}
	\caption{Sketch of the diabolo geometry, fixed boundary condition and prescribed displacement and rotation.}
	\label{fig:diabolo_sketch}
\end{figure}

The diabolo problem comes from \cite{Chagnon2006}. It is a simple body that combines a uniaxial tensile and a torsional load, labelled with $u$ and $\theta$ in Fig.~\ref{fig:diabolo_sketch}. In the original paper from which the example is taken, it is tested against several load combinations, but in this paper only the first load combination with $u = 30$ mm and $\theta = 5$ rad is taken. The lower base of the diabolo is fixed and the upper base is connected to the reference point via a rigid coupling.

\begin{figure}
	\centering
	\includegraphics{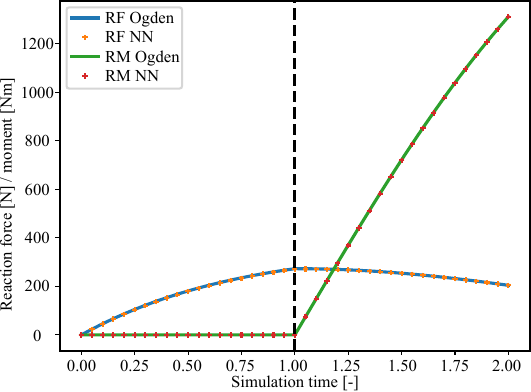}
	\caption{Reaction force and moment at the node where displacement and rotation are prescribed. The median relative error for the reaction force is $0.42\%$ and for the moment $0.32\%$. The dashed vertical line marks the end of uniaxial tension step and beginning of torsional load step.}
	\label{fig:diabolo_reactions}
\end{figure}

The diagram of the reaction forces and moments is shown in Fig.~\ref{fig:diabolo_reactions}. The reactions are taken at the reference point at which the displacement and rotation are prescribed. The NN model agrees with the reference model, with relative errors of about $0.4\%$.

\begin{figure}
	\centering
	\begin{subfigure}{\textwidth}
		\centering
		\includegraphics{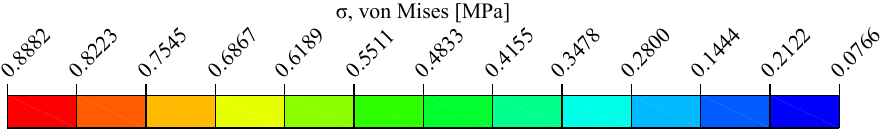}
	\end{subfigure}
	\begin{subfigure}{0.5\textwidth}
		\centering
		\includegraphics{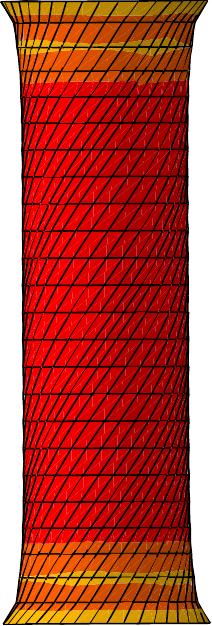}
		\caption{NN solution.}
	\end{subfigure}%
	\begin{subfigure}{0.5\textwidth}
		\centering
		\includegraphics{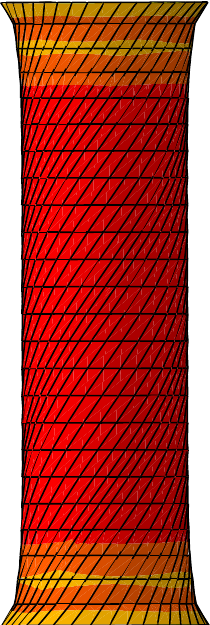}
		\caption{Ogden solution.}
	\end{subfigure}
	\caption{Von Mises stress plots of the diabolo problem using the proposed NN solution and the Ogden model with Mullins effect. The maximum/minimum stress values are 0.8864/0.0766 MPa and 0.8882/0.0769 MPa for the NN and Ogden solutions respectively.}
	\label{fig:diabolo_von_mises}
\end{figure}

The von Mises stress plots are shown superimposed on the deformed shape at the end of the simulation, in Fig.~\ref{fig:diabolo_von_mises}. The two plots are virtually indistinguishable with the relative error of maximum stresses of $0.01\%$ and of the minimum stresses of $0.48\%$.

\begin{figure}
	\centering
	\includegraphics{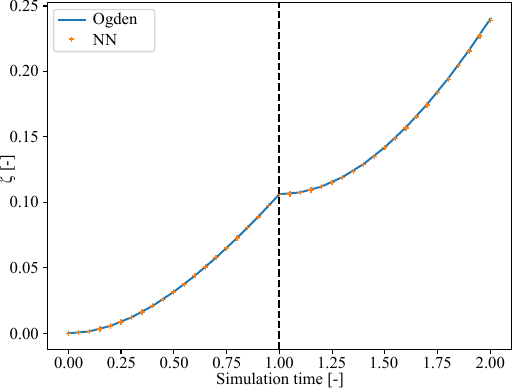}
	\caption{Evolution of the damage variable $\zeta$ during simulation. The vertical dashed line represents the end of the uniaxial tension step and beginning of the torsional load step.}
	\label{fig:diabolo_damage_evolution}
	
\end{figure}

\begin{figure}
	\centering
	\begin{subfigure}{\textwidth}
		\centering
		\includegraphics{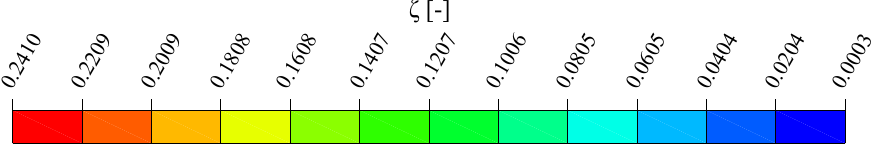}
	\end{subfigure}\par\bigskip
	\begin{subfigure}{0.5\textwidth}
		\centering
		\includegraphics{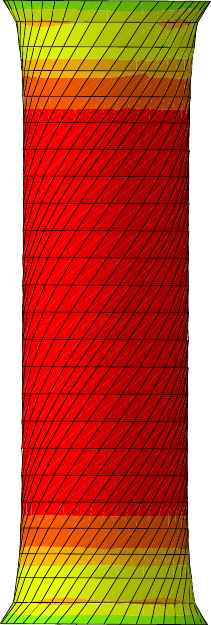}
		\caption{NN solution.}
	\end{subfigure}%
	\begin{subfigure}{0.5\textwidth}
		\centering
		\includegraphics{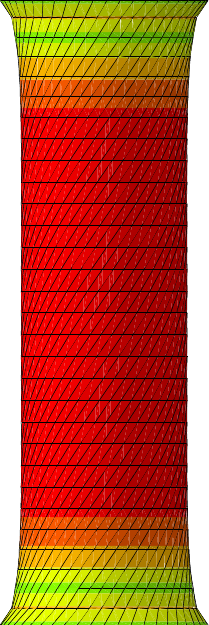}
		\caption{Ogden solution.}
	\end{subfigure}
	\caption{Plots of the damage variable values on the full diabolo geometry at the end of the simulation.}
	\label{fig:diabolo_full_damage_plot}
\end{figure}

The evolution of the damage parameter during the simulation at the centre node on the cylinder surface is given in Fig.~\ref{fig:diabolo_damage_evolution}. The NN solution is in good accordance with the reference Ogden solution and the median relative error is 0.39\%. In Fig.~\ref{fig:diabolo_full_damage_plot} the plot of the damage variable $\zeta$ on the entire geometry is shown and  the figures are in good agreement.

\begin{figure}
	\centering
	\includegraphics{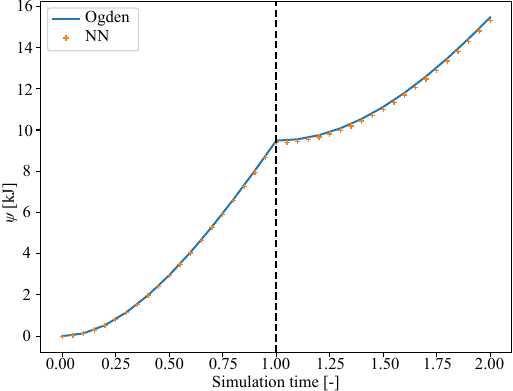}
	\caption{Evolution of the strain energy for the whole model during simulation. The vertical dashed line represents the end of the uniaxial tension step and beginning of the torsional load step.}
	\label{fig:diabolo_energy_whole_model}
\end{figure}

In Fig.~\ref{fig:diabolo_energy_whole_model} the plot of the total strain energy evolution during simulation iis given and it can be seen that the NN successfully captured the underlying behaviour. The median error along the entire simulation was 1.05\%.

\subsubsection{Diabolo - Mullins subnetwork}\label{sec:diabolo_subnet}

\begin{figure}
	\centering
	\begin{subfigure}{0.5\textwidth}
		\centering
		\includegraphics[width=\textwidth]{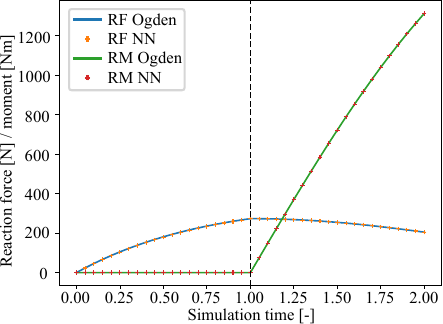}
		\caption{NN with subnetwork, $\beta$ omitted.}
		\label{fig:naive_diabolo_reactions}
	\end{subfigure}%
	\begin{subfigure}{0.5\textwidth}
		\centering
		\includegraphics[width=0.955\textwidth]{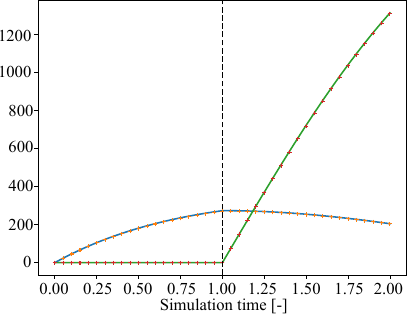}
		\caption{NN with subnetwork, $\beta$ is trainable.}
		\label{fig:non_naive_diabolo_reactions}
	\end{subfigure}
	\caption{Diabolo reaction force and moment at the node where the displacement and rotation were prescribed. Results shown for NNs with a subnetwork for modelling damage. The dashed vertical line marks the end of uniaxial tension step and beginning of torsional load step.}
	\label{fig:diabolo_reactions_subnet}
\end{figure}

\begin{figure}
	\centering
	\begin{subfigure}{0.5\textwidth}
		\centering
		\includegraphics[width=0.99\textwidth]{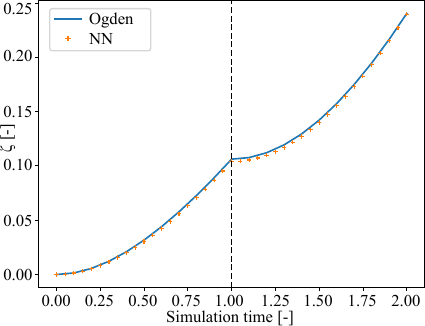}
		\caption{NN with subnetwork, $\beta$ omitted.}
		\label{fig:naive_diabolo_damage_evolution}
	\end{subfigure}%
	\begin{subfigure}{0.5\textwidth}
		\centering
		\includegraphics[width=0.99\textwidth]{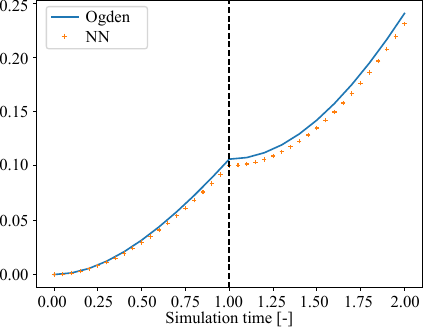}
		\caption{NN with subnetwork, $\beta$ is trainable.}
		\label{fig:non_naive_diabolo_damage_evolution}
	\end{subfigure}
	\caption{Evolution of the damage variable $\zeta$ during simulation. Results shown for NNs with a subnetwork for modelling damage. The dashed vertical line marks the end of uniaxial tension step and beginning of torsional load step.}
	\label{fig:diabolo_damage_evolution_subnet}
\end{figure}

In tIn this section, the reaction force and moment as well as the evolution of the damage variable on the diabolo are presented using NNs with a subnetwork for modelling the damage. Fig.~\ref{fig:diabolo_reactions_subnet} shows the evolution of the reaction force and moment. The median relative error of the reaction force and moment is 0.24\% and 0.15\% for the subnetwork with $\beta$ omitted, and 0.76\% and 0.65\% for the subnetwork where $\beta$ is a trainable parameter. The results show good agreement with the underlying Ogden model.
In Fig.~\ref{fig:diabolo_damage_evolution_subnet} the evolution of the damage variable $\zeta$ is shown for both the subnetwork approaches and here the results show an interesting behaviour. For simple uniaxial tension in Fig.~\ref{fig:non_naive_uniaxial_damage}, the damage variable was in agreement until higher damage values were reached, while in the diabolo example in Fig.~\ref{fig:non_naive_diabolo_damage_evolution}, the damage variable begins to be underestimated early on when using the approach with $\beta$ as a trainable parameter. In Fig.~\ref{fig:naive_diabolo_damage_evolution}, however, the damage variable is calculated correctly all the time and shows similar results to the NN with the specialised damage function.

\section{Conclusion}\label{sec:conclusion}
This paper presents a neural network model for Mullins type isotropic damage in hyperelastic materials. It is developed by taking into account certain conditions from solid mechanics such as \textit{objectivity, material symmetry, polyconvexity, thermodynamic consistency, normalisation} and \textit{non-negativity of energy}. The NN architecture itself is based on activation functions developed in accordance with conventional models of hyperelasticity. It has been further extended to use a simple conventional measure of damage without compromising any of the above conditions. It has also been demonstrated that the damage evolution can be successfully captured using NNs in place of conventional damage models while satisfying physical constraints of the damage parameter. Lastly, a modelling strategy is introduced where the NN block that models hyperelastic behaviour is reused in the same NN model allowing it to be trained directly on the damaged stresses.

In all examples presented, the NN model without the constraints of \textit{polyconvexity} and \textit{non-negativity of the energy} has been shown to be accurate, simple to develop with existing machine learning libraries and easy to use in commercial engineering software. Although trained on 3 simple deformation modes under plane stress conditions, it successfully captures the full hyperelastic behaviour in 3 dimensions. Also, it has been demonstrated that overconstraining the NNs to be polyconvex is too restrictive since they were not able to replicate the underlying material behaviour.

The approach presented here can be extended for application to more complex damage phenomena, with the subnetwork approaches offering great flexibility allowing the use of general NNs for modelling behaviours without explicitly assuming their form while fulfilling physical restraints. A natural progression would be to modify the NN model for use in damage with residual strains, adopting the general modelling techniques from existing models. Another possibility would be to use the current NN model for anisotropic behaviour in conjunction with damage by adding the appropriate pseudo-invariants.

In summary, a general NN hyperelastic material model that takes into consideration the Mullins effect has been developed that satisfies certain conditions from solid mechanics and has the potential to capture any desired hyperelastic behaviour while maintaining a simple form that is not much larger than conventional polynomial models.



\FloatBarrier

\section*{Acknowledgements}

This work was supported in part by the University of Rijeka under project number uniri-iskusni-tehnic-23-37 and in part by the Croatian Science Foundation under project IP-2019-04-4703. All the support is gratefully acknowledged.

\section*{CRediT authorship contribution statement}

\textbf{Martin Zlatić:} Conceptualization, Methodology, Software, Validation, Formal analysis, Visualization, Writing - Original Draft, Writing - Review \& Editing. \textbf{Marko Čanađija:} Resources, Writing - Review \& Editing, Funding Acquisition, Supervision

\section*{Declaration of competing interest}

The authors declare that they have no known competing financial or personal relationships that could have appeared to influence the work reported in this paper.

\appendix

\section{Abaqus implementation details}\label{sec:implementation_details}

The NN model was implemented into Abaqus through the UHYPER subroutine. It requires the partial derivatives of the strain energy $\psi_\text{NN}$ with respect to the invariants. The derivatives can be given in explicit form, following the NN architecture given in Section~\ref{sec:NN_Mullins}.

\begin{equation}
	\frac{\partial{\psi_\text{NN}}}{\partial{I_k}} = (1-\zeta)\frac{\partial{\psi_0}}{\partial{I_k}}, \quad
	\frac{\partial^2{\psi}}{\partial{I_k}\partial{I_l}} = (1-\zeta)\frac{\partial^2{\psi_0}}{\partial{I_k}\partial{I_l}},
\end{equation}
with $k,l = 1,2$. The partial derivatives of $\psi_0$ given as
\begin{equation}
	\frac{\partial{\psi_0}}{\partial{I_k}} = \sum_{i=1}^{n}w_{3,i}\frac{\partial{g_i}}{\partial{I_k}}, \quad
	\frac{\partial^2{\psi_0}}{\partial{I_k}\partial{I_l}} = \sum_{i=1}^{n}w_{3,i}\frac{\partial^2{g_i}}{\partial{I_k}\partial{I_l}}.
\end{equation}
The $1^\text{st}$ order derivatives are as follows:
\begin{equation}
	\frac{\partial{g_i}}{\partial{I_k}} = w_{k,i}\alpha_i\underbrace{\exp\big[\alpha[w_{1,i}(I_1-3) + w_{2,i}(I_2 - 3)]\big]}_{C_i},
	\label{eq:1st_order_act_func_derivative}
\end{equation}
and the $2^\text{nd}$ order derivatives are given as:
\begin{equation}
	\frac{\partial^2{g_i}}{\partial{I_k}\partial{I_l}} = w_{k,i}w_{l,i}\alpha_i^2\exp\big[\alpha_i[w_{1,i}(I_1-3) + w_{2,i}(I_2 - 3)]\big], \quad \text{for } k,l = 1,2; i = 1,...,n.
	\label{eq:activation_2nd_derivative_generic} 
\end{equation}

The outline of the procedure is given in Algorithm~\ref{algo:UHYPER_procedure}. Note that the expressions in lines \ref{alg_h1}, \ref{alg_h2}, \ref{alg_h3} match the expression for the activation function from Eq.~\eqref{eq:linear_exponential}, the expression in line \ref{alg_ci} matches the one given in  Eq.~\eqref{eq:1st_order_act_func_derivative}. The damage parameter is calculated in line \ref{alg_1-zeta} according to Eq.~\eqref{eq:1-zeta_expanded} when using the specialised damage function inspired by the CANN approach.

When implementing a NN with a subnetwork that models the damage in place of the specialized function presented in Eq.~\eqref{eq:1-zeta_expanded} the procedure outlined in Algorithm~\ref{algo:UHYPER_procedure}. The only thing that needs to be replaced in Algorithm~\ref{algo:UHYPER_procedure} is line \ref{alg_1-zeta} with the code snippet given in Algorithm~\ref{algo:subnetwork_snippet}.

\noindent\textbf{Remark 1} If the subnetwork without the trainable parameter $\beta$ is implemented, then it is possible that for high values of $\psi_\text{0,max}$ the damage $\zeta$ is equal to 1 and in Algorithm~\ref{algo:subnetwork_snippet} in line \ref{alg_1-zeta_subnet} the value can be 0 thus rendering all the derivatives 0. This does not occur when simple tests like uniaxial, equibiaxial or planar tension are calculated but it does occur in complex loading scenarios such as the solid disc example. To avoid this the variable W6 should be set to nearly 1 (e.g. 0.9999).  

\noindent\textbf{Remark 2} If an NN with $\beta$ is trained it is possible that it trains to mimick the situation from Remark 1 so that $\beta$ is trained to be nearly 1 (i.e. 0.99999 or similar) and the NN behaves exactly like the one where $\beta$ is omitted, i.e. it correctly predicts the Mullins energy, the undeformed energy and the damage variable, as in Fig.~\ref{fig:subnet_comparison}. This is however a niche situation and in most cases $\beta$ is less than 1 like in Fig.~\ref{fig:subnet_comparison}, so it is not necessarily reproducible. However, training without $\beta$ always yields almost the same result (a small variance exists due to the random initialization of the weights) and is therefore preferred. 

\begin{algorithm}
	\caption{UHYPER procedure.}
	\label{algo:UHYPER_procedure}
	\begin{algorithmic}[1]
		\State PSIM = 0
		\Statex \textbf{Calculating undamaged strain energies.}
		\State DO J = 1, NNEUR \Comment{NNEUR is the number of neurons.}
		\State $\quad$ H = EXP(ALPHA(J)*(W11(J)*(I1MAX-3)+W21(J)*(I2MAX-3))) -1\label{alg_h1}
		\State $\quad$ PSIM = PSIM + W3(J)*H \Comment{Maximum energy.}
		\State $\quad$ H = EXP(ALPHA(J)*(W11(J)*(BI1-3)+W21(J)*(BI2-3))) - 1\label{alg_h2}
		\State $\quad$ DUDI(1) = DUDI(1) + W3(J)*H \Comment{Current energy.}
		\State END DO
		\Statex \textbf{Check for new maximum energy.}
		\State IF (DUDI(1).GT.PSIM) THEN
		\State $\quad$ I1MAX = BI1
		\State $\quad$ I2MAX = BI2
		\State $\quad$ PSIM = DUDI(1)
		\State $\quad$ STATEV(1) = I1MAX \Comment{User variable.}
		\State $\quad$ STATEV(2) = I2MAX \Comment{User variable.}
		\State END IF
		\Statex \textbf{Calculate damage parameter.}
		\State A = 1D0-ZETA\_MAX+ZETA\_MAX*EXP(-PSIM/IOTA)\label{alg_1-zeta}
		\Statex \textbf{Calculate the derivatives.}
		\State DO I = 1, NNEUR
		\State $\quad$ H = EXP(ALPHA(I)*(W11(I)*(BI1-3)+W21(I)*(BI2-3))) - 1\label{alg_h3}
		\State $\quad$ CI = EXP(ALPHA(I)*(W11(I)*(BI1-3)+W21(I)*(BI2-3)))\label{alg_ci}
		\State $\quad$ DUDI(2) = DUDI(2) + W3(I)*ALPHA(I)*W11(I)*CI
		\State $\quad$ DUDI(3) = DUDI(3) + W3(I)*ALPHA(I)*W21(I)*CI
		\State $\quad$ DUDI(4) = DUDI(4) + W3(I)*ALPHA(I)**2*W11(I)**2*CI
		\State $\quad$ DUDI(5) = DUDI(5) + W3(I)*ALPHA(I)**2*W21(I)**2*CI
		\State $\quad$ DUDI(6) = DUDI(6) + W3(I)*ALPHA(I)**2*W11(I)*W21(I)*CI
		\State END DO
		\Statex \textbf{Apply damage.}
		\State DO I = 1, 6
		\State $\quad$ DUDI(I) = A*DUDI(I)
		\State END DO
	\end{algorithmic}
	
\end{algorithm}
\FloatBarrier

\begin{algorithm}
	\caption{Subnetwork snippet.}
	\label{algo:subnetwork_snippet}
	\begin{algorithmic}[1]
		\Statex \textbf{Subnetwork is used to calculate energy from PSIM.}
		\State AUX = 0D0
		\State DO I = 1, 5
		\State $\quad$ AUX = AUX + (EXP(ALPHA2(I)*W4(I)*PSIM)-1)*W5(I) 
		\Statex \Comment{W4, ALPHA2, W5 are the subnetwork weights.}
	  	\State END DO
	  	\State ZETA = W6*TANH(AUX) \Comment{W6 is either $\beta$ or 0.9999.}
	  	\State A = 1D0 - ZETA \label{alg_1-zeta_subnet}
	\end{algorithmic}

\end{algorithm}


\end{document}